\documentclass[traditabstract]{aa}  
\usepackage{graphicx}
\usepackage{amsmath}
\usepackage[colorlinks=true,urlcolor=blue,citecolor=blue,linkcolor=blue]{hyperref}
 \usepackage[varg]{txfonts}
\usepackage{subfigure}
\usepackage{url}
\usepackage{natbib}
\bibpunct{(}{)}{;}{a}{}{,}
\usepackage[usenames]{color}
\usepackage{amssymb}
\def\mathbi#1{\textbf{\em #1}}

\begin{document}
\frenchspacing
\title{Physical Properties of a Sunspot Chromosphere with Umbral Flashes}
\author{
  J. de la Cruz Rodr\'iguez\inst{1},
 L. Rouppe van der Voort\inst{2},
 H. Socas-Navarro\inst{3}
\and  M. van Noort\inst{4}
}
\authorrunning{de la Cruz Rodr\'iguez et al.}

\institute{  
  Department of Physics and Astronomy, Uppsala University, 
  Box 516, SE-75120 Uppsala, Sweden
  \and
  Institute of Theoretical Astrophysics, University of Oslo, 
  P.O. Box 1029 Blindern, N-0315 Oslo, Norway
  \and
   Instituto de Astrof\'isica de Canarias, Avda V\'ia L\'actea S/N, 
  La Laguna 38205, Tenerife, Spain
  \and
  Max-Planck Institute for Solar System Research, Max-Planck Straße 1, 
  37412 Katlenburg-Lindau, Germany
}


\abstract {We present new high-resolution spectro-polarimetric \ion{Ca}{ii}~$\lambda8542$
  observations of umbral flashes in sunspots. At nearly $0\farcs18$,
  and spanning about one hour of continuous observation, this is the
  most detailed dataset published thus far. Our study involves both
  LTE and non-LTE inversions (but includes also a {\em weak field}
  analysis as a sanity check) to quantify temperatures, mass flows and
  the full magnetic field vector geometry. We confirm earlier reports
  that UFs have very fine structure with hot and cool material
  intermixed at sub-arcsecond scales. The shock front is roughly
  1,000~K hotter than the surrounding material. We do not observe
  significant fluctuations of the field in the umbra. In the penumbra,
  however, the passage of the running penumbral waves alter the
  magnetic field strength by some 200~G (peak-to-peak amplitude) but
  it does not change the field orientation (at least not significantly
  within our sensitivity of a few degrees). We find a trend of decreasing high-frequency modulation power for more inclined magnetic fields for the line-of-sight velocity and magnetic field strength. In the outer penumbra we find an absence of high frequency power while there is increasingly more power at high frequencies towards the umbra. }

\maketitle

\section{Introduction}\label{sec:intro}

Umbral flashes (UFs) are sudden brightenings observed in the
chromospheric core of the \ion{Ca}{ii}\ lines. They tend to appear
with a periodicity of roughly 3 minutes \citep{BT69, W69} and it is
now reasonably well established that they are the chromospheric
counterpart of the photospheric oscillation \citep{KMU81,T84}. The
waves, propagating from the underlying photosphere, steepen and become
non-linear as the atmospheric density drops by several orders of
magnitude. Eventually, the waves develop into shocks which are
observable as blue-shifted emission reversals in the \ion{Ca}{ii} line
cores. A semi-empirical study by
\citet{2011felipe} allowed to reproduce multi-line sunspot
observations using a 3D non-linear MHD code \citep{2010felipe}. A more
theoretical study by \citet{2010bard} analyzed which 
modes contribute to the
formation of UFs and investigated the role of the coronal radiation in the
strength of these shock events. 

The development of spectro-polarimetry has paved the way for new
advances in our understanding of UFs, mostly aided by the
characteristic nearly anti-symmetric shape of the Stokes~$V$
profile. The first polarimetric observations of UFs
\citep{SNTBRC00c,SNTBRC00b} already revealed the occurrence of
``anomalous'' Stokes~$V$ profiles during the flash events. The
anomalous profiles could only be explained as the superposition of two
regular profiles of opposite polarity and considerably Doppler-shifted
with respect to each other. The obvious conclusion then is that, while
UFs are a large-scale phenomenon, forming emission patches that are
several arc-seconds across, they harbor fine-scale
structure. Therefore, even inside a single pixel, one has a mixture in
the horizontal direction where a certain filling factor is occupied
by a ``quiet'' component (no reversal and zero or slightly downflowing
velocity) while the rest is occupied by the shock-wave (line core
emission reversal and strong upflows).

Additional indirect evidence supporting the two-component scenario
described above was presented by \cite{CSNc+05} using observations of
the \ion{He}{i}\ 10830\ \AA\ multiplet acquired with the TIP
instrument \citep{CRHBR+99}. However, such fine structure inside the
umbral flash patches was not directly seen until years later, with the
analysis of \cite{SNMC+09} using high-resolution images of the
\ion{Ca}{ii}\ H line core from the Hinode satellite \citep{Hinode1,
  Hinode5}. With the increased spatial resolution offered today by
some ground-based telescopes, it has been possible to confirm this
finding by other authors \citep{BHS13, thesis-vasco}.

Umbral flashes seem to be connected to the phenomenon known as running
penumbral waves, an oscillatory pattern originally observed in
H$\alpha$ images (but now also in \ion{He}{i}, as shown by
\citealt{BLS07}, and \ion{Ca}{ii} in this work). The wave pattern is
seen to move horizontally across the penumbra, running away from the
umbra in the radial direction \citep{TAM00, RvdVRS+03}. The
relationship between UFs and penumbral waves is not only suggested by
visual appearance. The first spectro-polarimetric observations of
\cite{LASNM01} showed evidence that the shock bends radially outwards
following the field lines as it propagates upwards. More recently, the
inversions of infrared Stokes spectra carried out by \cite{BLS07}
indicate that the same wave propagation that gives rise to the umbral
flashes is at work in the penumbra and produces the running penumbral
waves in a scenario where the magnetic field is more inclined.

The paper is arranged as follows: the observations and data processing
are described in Sect.~\ref{sec:obs}. A detail description of the
umbral flashes in our observations and a simple analysis of the
magnetic field strength, using the weak-field approximation are
provided in Sect.~\ref{sec:uf}. In Sect.~\ref{sec:inv} we describe the
non-LTE inversion code \textsc{Nicole} and our results are presented
in Sect.~\ref{sec:aum} and Sect.~\ref{sec:rpw}. Finally, we summarize
our conclusions in Sect.~\ref{sec:disc}.

\section{Observations and data processing}\label{sec:obs}
The data used for this work were acquired at the Swedish 1-m Solar
Telescope \citep[SST,][]{2003scharmer} using the CRisp Imaging
SpectroPolarimeter \citep[CRISP,][]{2006scharmer,2008scharmer}.  CRISP
is a Fabry-Perot interferometer (FPI) consisting of two etalons
mounted in tandem. With this kind of instrumentation one is able to
record quasi-monochromatic images at a given wavelength of choice. A
line profile may then be constructed by scanning the corresponding
spectral range, obtaining images at each line position.
 
The observations were obtained on 2011 May 4 at 08:30~UT. 
The target
was a sunspot (AR11204) at heliocentric $(x,y)=(-336\arcsec,332\arcsec)$, observing angle $30\degr$ ($\mu=0.87$).

The dataset includes full-Stokes observations in
\ion{Ca}{ii}\ $\lambda8542$ and non-polarimetric images in
\ion{H}{i}\ $\lambda6563$ ($H\alpha$).
The \ion{Ca}{ii}\ $\lambda8542$
line was observed at 14 line positions with a sampling of 75~m\AA
\ between $\Delta\lambda=\pm300$~m\AA \ (an effective integration time
of 900~ms), and points at $\Delta\lambda=\pm400, \pm860, +3100$~m\AA
\ from line center (with an effective integration time of 210~ms).
Images are acquired concatenating four polarization states. Each
polarization state is a linear combination of the for Stokes
parameters $(I,Q,U,V)$. The noise, relative
to the continuum intensity, is $\sigma(Q,U,V)=(0.09, 1.0,1.3)\times10^{-3}$.
The $H\alpha$ line is observed at
$\Delta\lambda=+0, +1300$~m\AA \ from line center.

\begin{figure}[]
      \centering
        \resizebox{\hsize}{!}{\includegraphics[trim=0cm 0.0cm 0cm 0.0cm,
          clip]{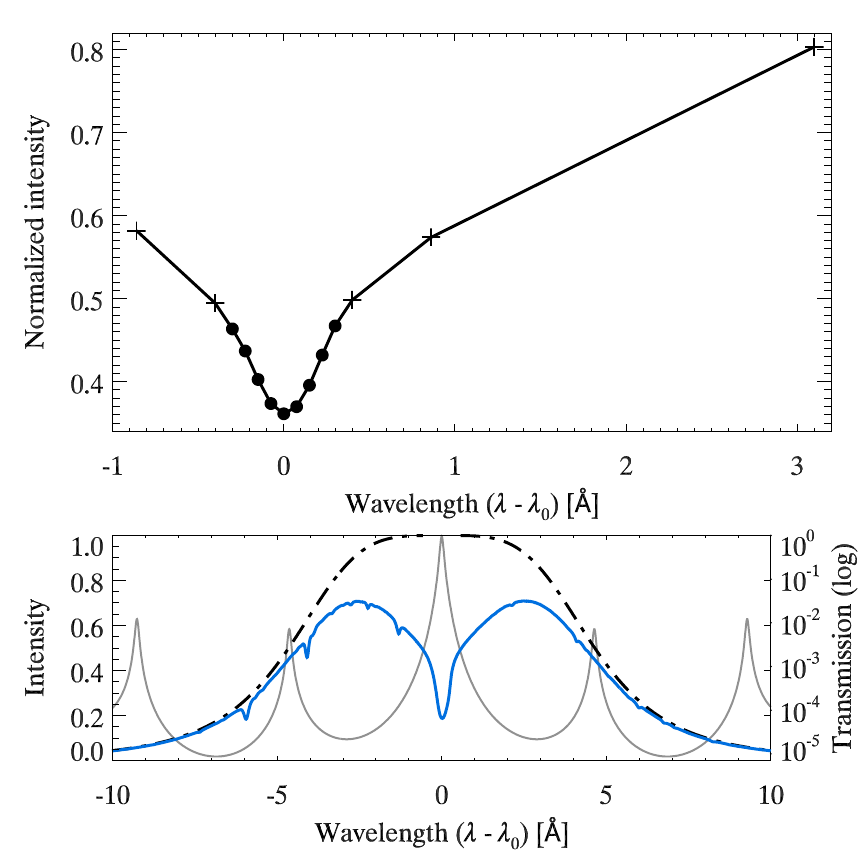}}
       \caption{The \ion{Ca}{ii}~$\lambda8542$ observations were acquired at the line positions
         indicated on the line profile in the top panel. The exposure
         time is 900~ms (circles) and 210~ms (crosses). The bottom
         panel illustrates the CRISP instrumental profile at 854.0~nm
         (solid gray) in logarithmic scale. A solar-atlas profile
         multiplied with the prefilter (dashed-dotted line) is
         indicated in blue.}
        \label{fig:f1}
\end{figure}

Fig.~\ref{fig:f1} illustrates the observed line positions at
$\lambda8542$, which are marked on a quiet-Sun spatial average from
the surroundings of the sunspot. Ideally, one point in the continuum
would be desirable but the presence of secondary transmission lobes in
the CRISP transmission profile
restricts the observations to the central part of the prefilter (see
the bottom panel in Fig.~\ref{fig:f1}).

\begin{figure*}[]
      \centering
        \resizebox{\hsize}{!}{\includegraphics[trim=0cm 0.9cm 0cm 0.05cm,
          clip]{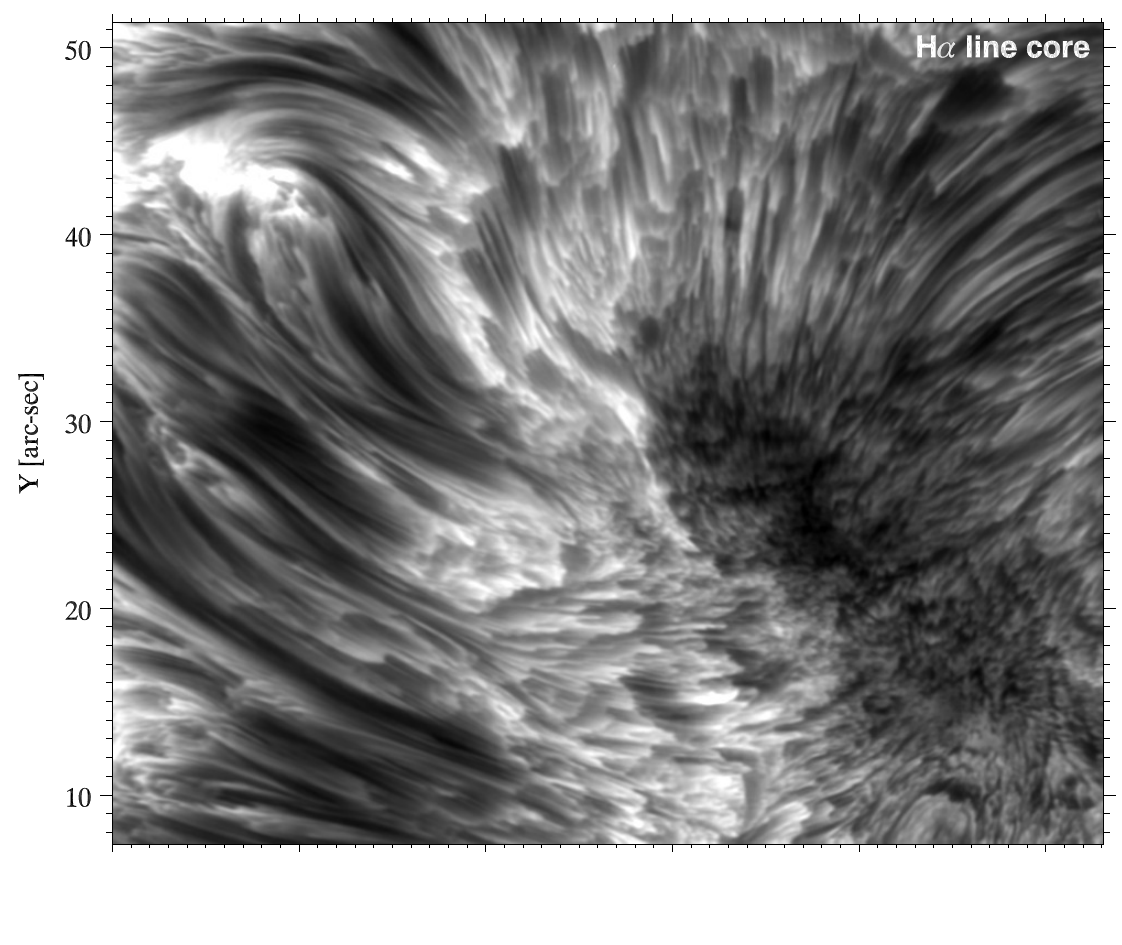}\includegraphics[trim=1cm 0.9cm 0cm 0.05cm,
          clip]{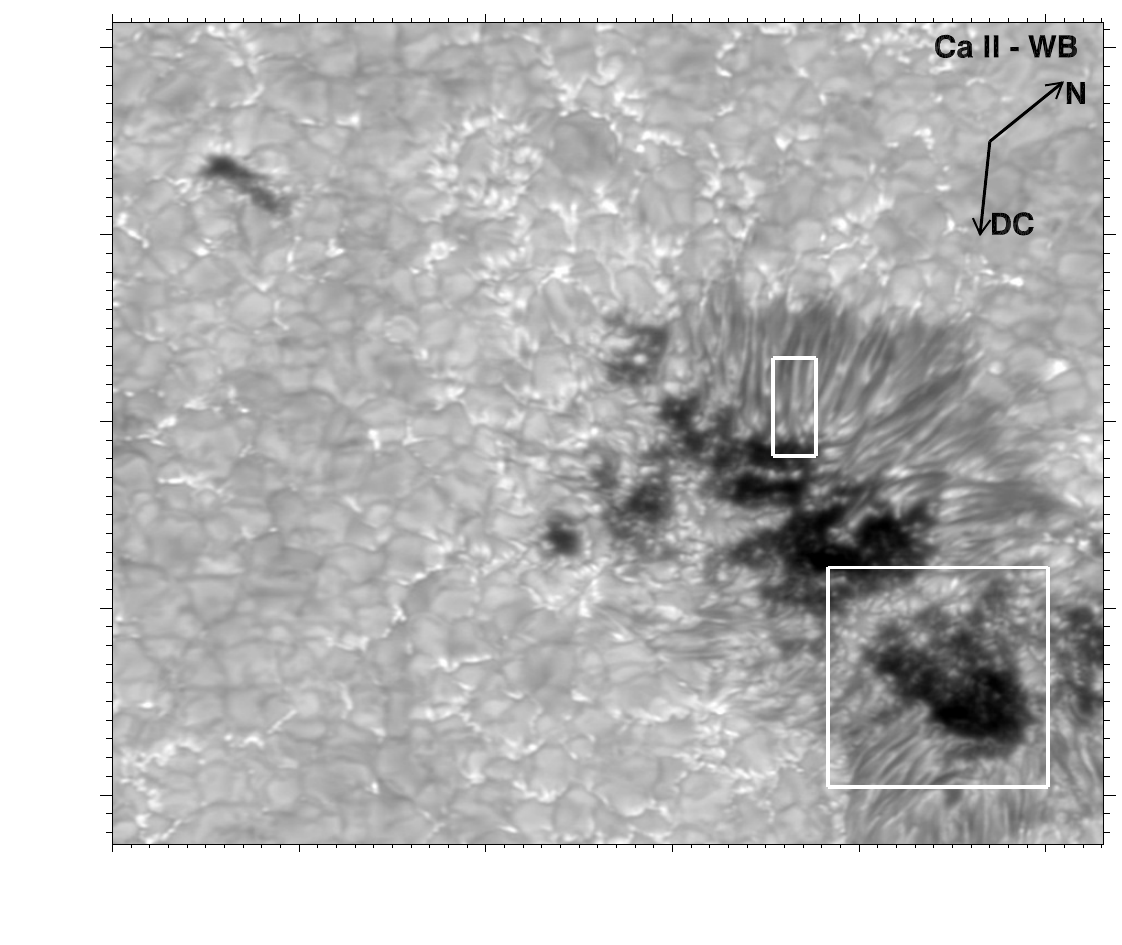}}
        \resizebox{\hsize}{!}{\includegraphics[trim=0cm 0.9cm 0cm 0.1cm,
          clip]{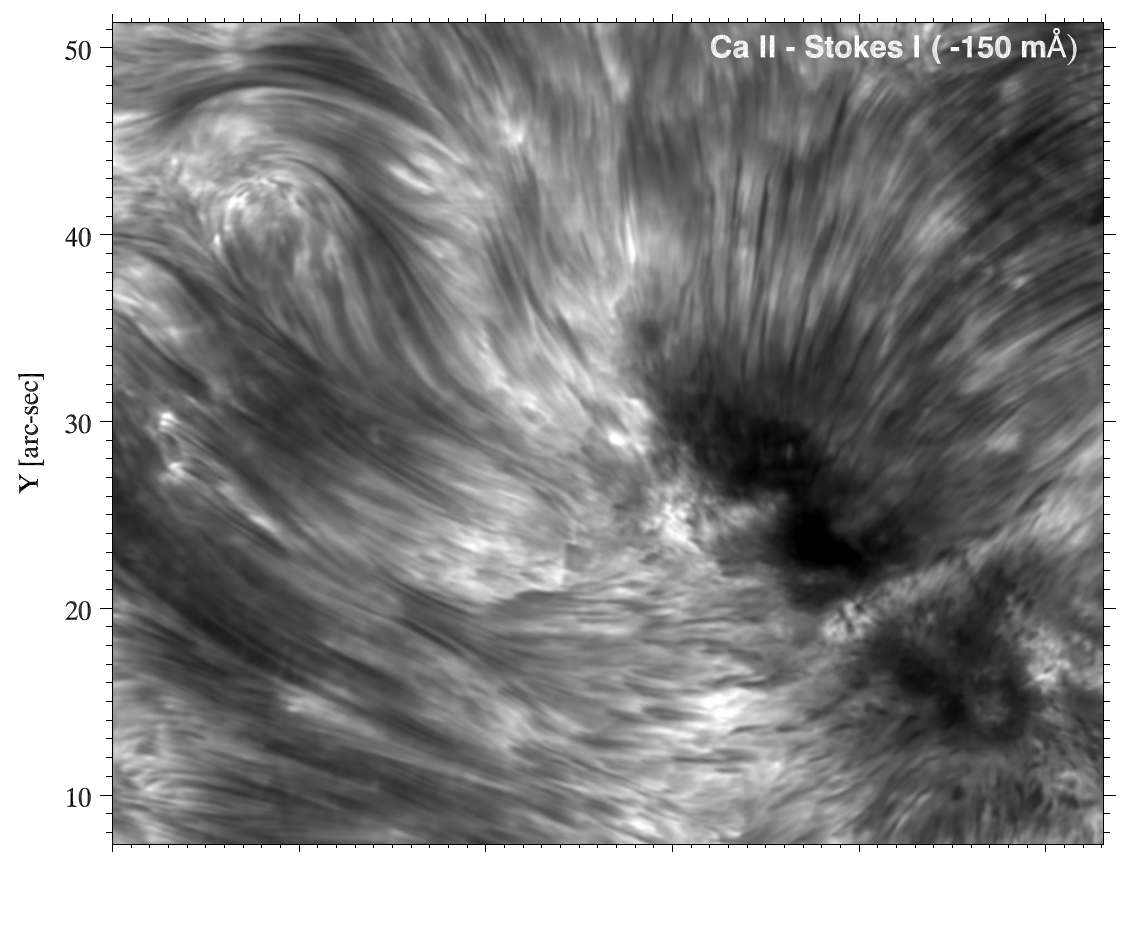}\includegraphics[trim=1cm 0.9cm 0cm 0.05cm,
          clip]{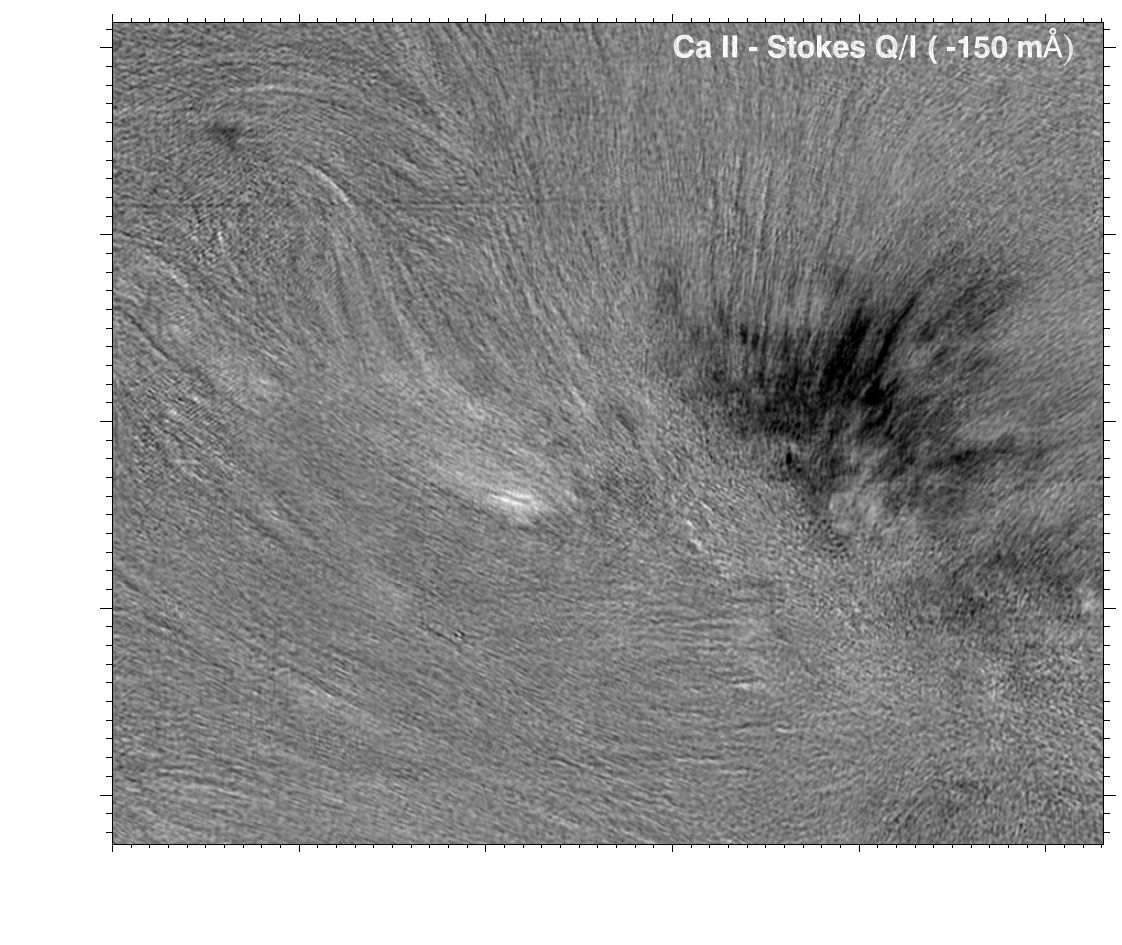}}
        \resizebox{\hsize}{!}{\includegraphics[trim=0cm 0.0cm 0cm 0.05cm,
          clip]{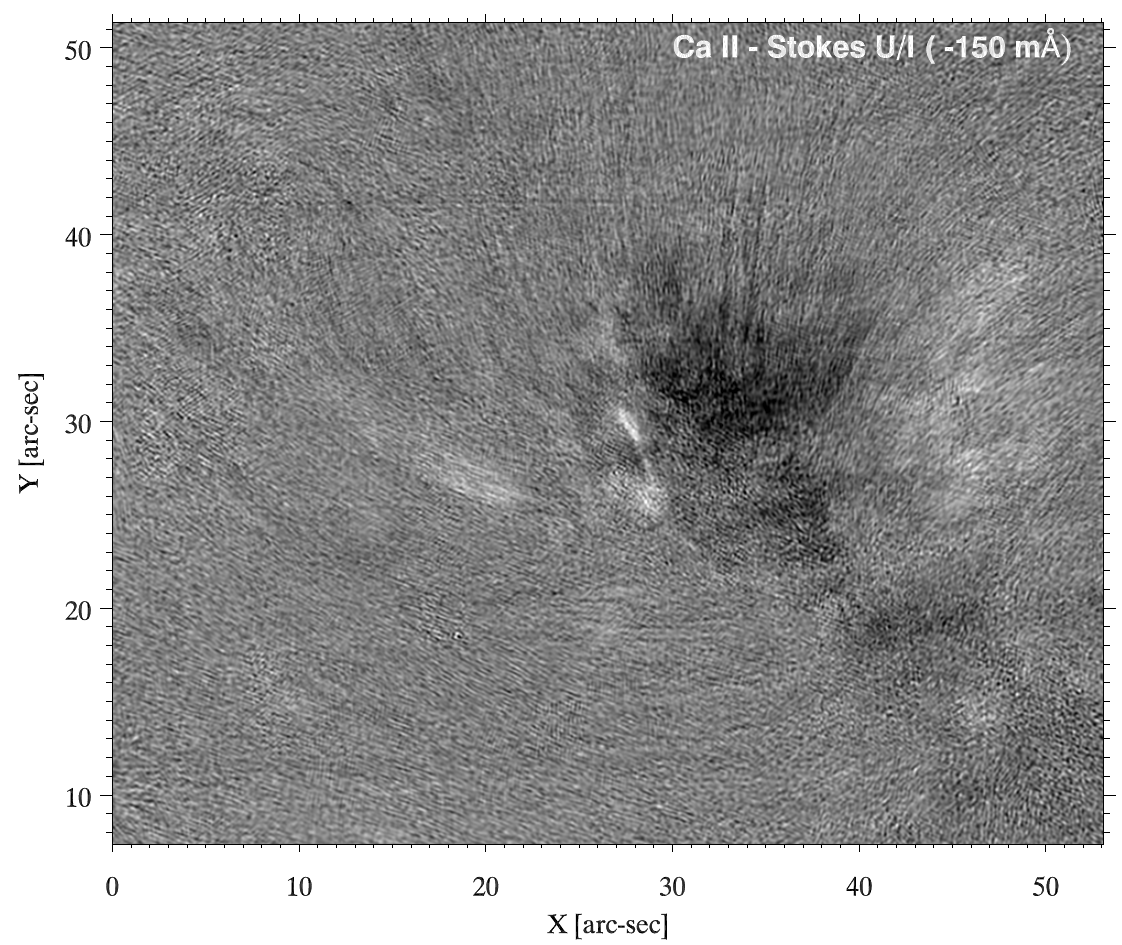}\includegraphics[trim=1cm 0.0cm 0cm 0.05cm,
          clip]{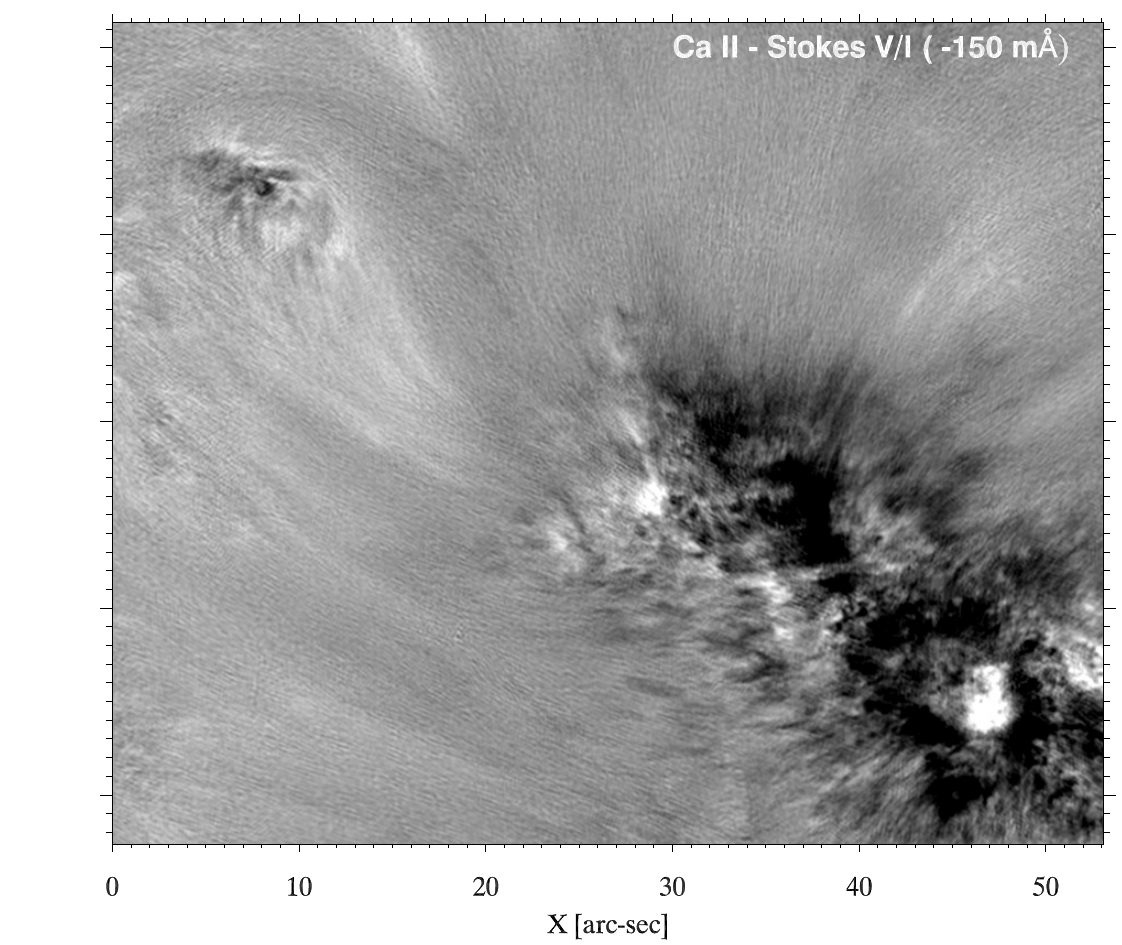}}

       \caption{CRISP filtergrams at the core of the $H\alpha$ line
         (top left). All other images are in
         \ion{Ca}{ii}~$\lambda8542$. A wide-band image showing the
         photosphere is displayed in the top right panel. The
         lowermost four panels show the chromosphere at
         $\Delta\lambda=-150$~m\AA \ from line core in Stokes~$I$
         (middle left), $Q/I$ (middle right), $U/I$ (bottom left) and
         $V/I$ (bottom right). The arrows on the wide-band image
         indicate the direction to the solar North and to disk center,
         respectively. The largest white box indicated in the
         wide-band image highlights the subfield that has been used for
         the non-LTE inversions, whereas the smaller square indicates
         a subfield used to study running penumbral waves.}
        \label{fig:f2}
\end{figure*}

The data processing was done following a similar approach as presented by
\cite{2011schnerr} but with some differences in the flat-fielding
process, further discussed in Appendix~\ref{sec:descatter}. High
spatial resolution and precise alignment between the sequentially
recorded images for different tunings was achieved with the
Multi-Object-Multi-Frame-Blind-Deconvolution
\citep[MOMFBD,][]{2005vannoort} technique for image restoration.

In certain areas, where the scale of the seeing motions is smaller
than the assumed size of the iso-planatic patch in the MOMFBD code,
residual distortions can be present within the images of a same scan,
or even the different polarization states at a given
wavelength. Reducing the patch size during the reconstruction can help
to minimize this effect, but in some cases it does not remove it entirely. \cite{2012henriques,thesis-vasco} proposed to include
extra wide-band images (one per wavelength and polarization state)
that do not contribute to the wavefront-sensing in the MOMFBD process,
but contain the same residual distortions as the narrow-band
images. The distortions are measured by comparing those extra images
with the anchor wide-band image and 
each narrow-band image is de-stretched accordingly. 

The polarimetric calibration of the instrument was performed in 2D,
resulting in a different Mueller matrix for each pixel \citep{2008vannoort}. The
telescope-induced polarization was removed using the SST model
developed by \citet{2010selbing} with the parameters fitted at
$\lambda8542$. The telescope model is further described in
Appendix~\ref{sec:telescope} below.

The time series has been compensated for telescope rotation, alignment
over time and residual rubber-sheet deformations 
\citep{1994shine}
using the wide-band images as reference. 

In order to normalize the data to a spatially-averaged quiet-Sun
continuum, we computed the intensity ratio between $\mu=0.87$ and
$\mu=1.0$ at $\Delta\lambda=+3100$~m\AA \ from line center, using
synthetic profiles, computed in non-LTE, from a 3D hydrodynamical
simulation \citep{2011delacruz}. This ratio was then applied to the
intensity at the same wavelength of the McMath-Pierce Fourier Transform
Spectrometer atlas \citep[hereafter FTS atlas,][]{fts-atlas}, which
has been previously convolved with the CRISP transmission profile. In
this manner we impose that the spatially-averaged quiet-Sun profile at
$\Delta\lambda=3100$~m\AA \ $(\mu=0.87)$, must have an intensity of
$0.792$ relative to the continuum at $\mu=1$.

Fig.~\ref{fig:f2} summarizes our observations. In the photosphere (see
the wide-band image in the top-right panel), photospheric granulation
surrounds an irregular sunspot. It has developed penumbra in
approximately $3/4$ of its circumference
and a light-bridge divides the
umbra in two halves (umbral cores). There is also a small pore on the
top-left corner of the image.

The $H\alpha$ chromospheric image (top-left) is dominated by very
dynamic fibrils which extend outwards, almost radially from the
umbra-penumbra boundary. Some fibrils originate in the fore-mentioned
pore and are oriented towards the sunspot, suggesting some magnetic
connection between the two.

The lowermost four panels correspond to the four Stokes parameters at
$-150$~m\AA \ from line center at $\lambda8542$. The Stokes~$I$ image
is remarkably similar to that in the core of $H\alpha$.

The Stokes~$Q$ and $U$ signals, produced by magnetic fields
perpendicular to the our line of sight, are mostly observed on the
limb side of the sunspot, preferentially along elongated fibrils that
extend in the
\emph{superpenumbra} \citep{1974bray}. The Stokes~$V$ image
(associated with the longitudinal magnetic field component) shows very
strong dynamics, with frequent sign changes due to emission reversals.

The time series shows that UFs and running penumbral waves are present
in the sunspot on both sides of the light-bridge.

\section{Umbral flashes at $0 \farcs 18$}\label{sec:uf}

\begin{figure*}[]
      \centering
        \resizebox{\hsize}{!}{\includegraphics[trim=0cm 0cm 0cm 0cm,
          clip]{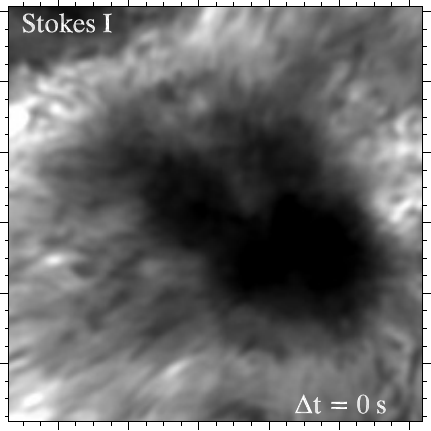}\includegraphics[trim=0cm 0cm 0cm 0cm,
          clip]{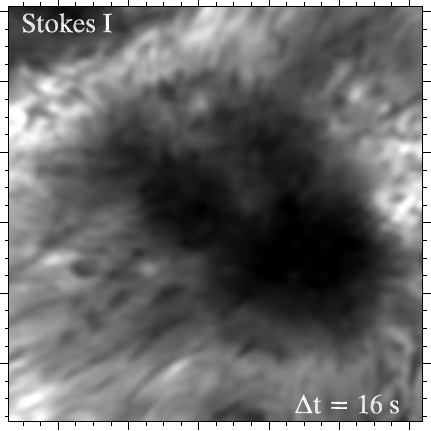}\includegraphics[trim=0cm 0cm 0cm 0cm,
          clip]{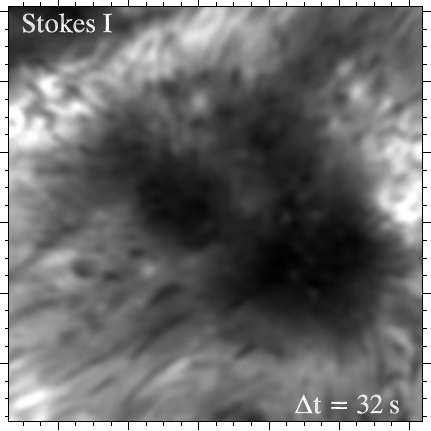}\includegraphics[trim=0cm 0cm 0cm 0cm,
          clip]{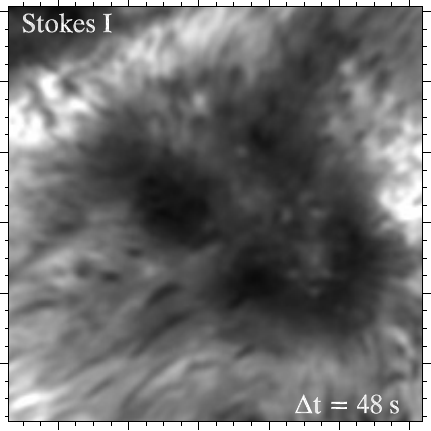}}
        \resizebox{\hsize}{!}{\includegraphics[trim=0cm 0cm 0cm 0cm,
          clip]{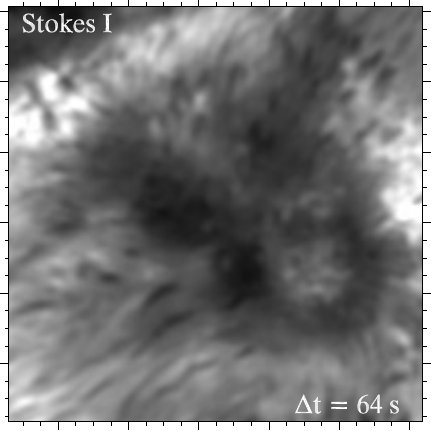}\includegraphics[trim=0cm 0cm 0cm 0cm,
          clip]{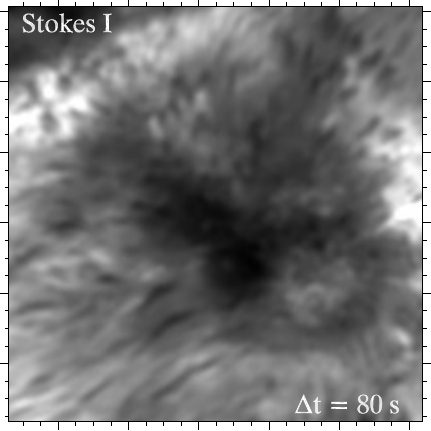}\includegraphics[trim=0cm 0cm 0cm 0cm,
          clip]{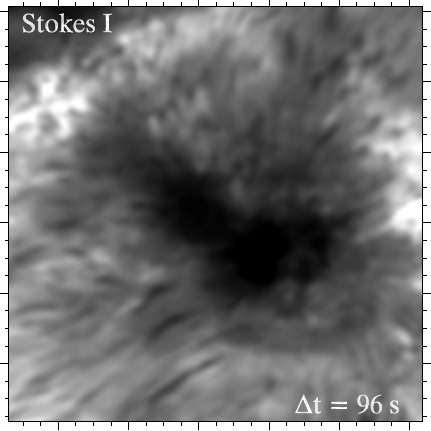}\includegraphics[trim=0cm 0cm 0cm 0cm,
          clip]{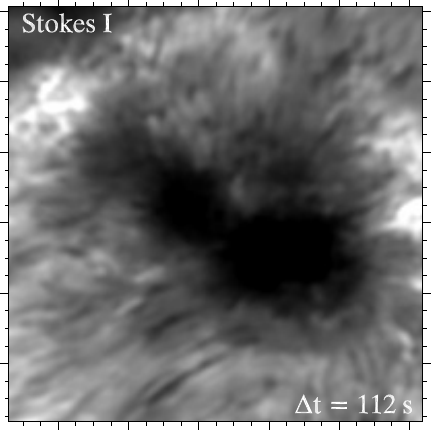}}
        \resizebox{\hsize}{!}{\includegraphics[trim=0cm 0cm 0cm 0cm,
          clip]{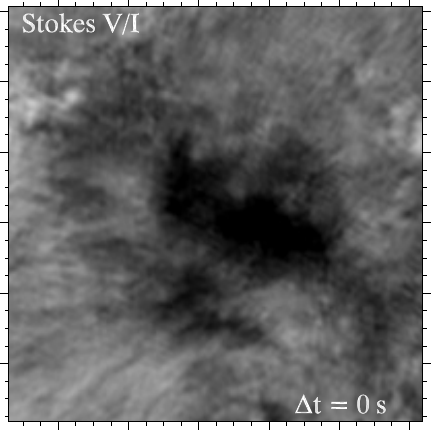}\includegraphics[trim=0cm 0cm 0cm 0cm,
          clip]{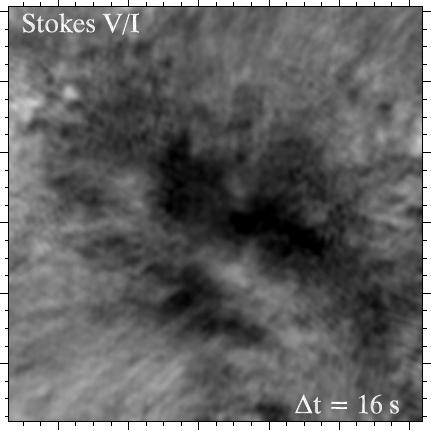}\includegraphics[trim=0cm 0cm 0cm 0cm,
          clip]{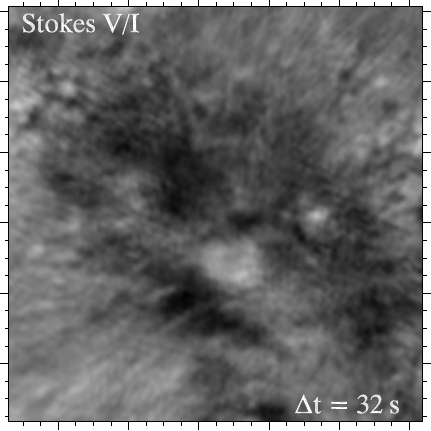}\includegraphics[trim=0cm 0cm 0cm 0cm,
          clip]{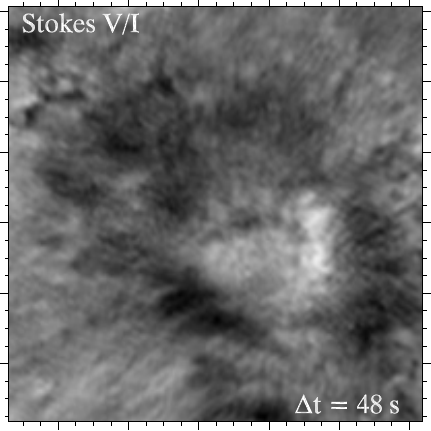}}
        \resizebox{\hsize}{!}{\includegraphics[trim=0cm 0cm 0cm 0cm,
          clip]{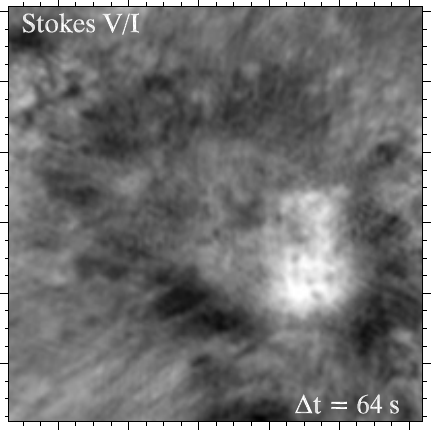}\includegraphics[trim=0cm 0cm 0cm 0cm,
          clip]{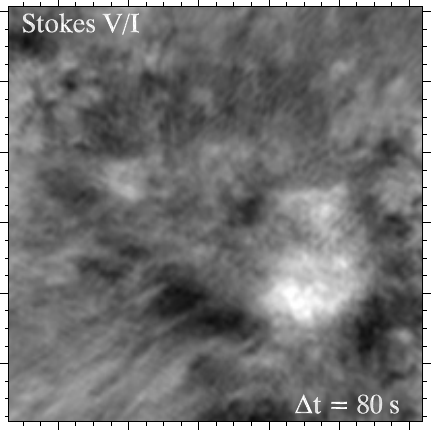}\includegraphics[trim=0cm 0cm 0cm 0cm,
          clip]{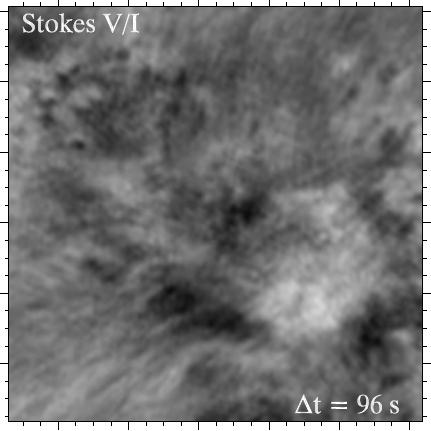}\includegraphics[trim=0cm 0cm 0cm 0cm,
          clip]{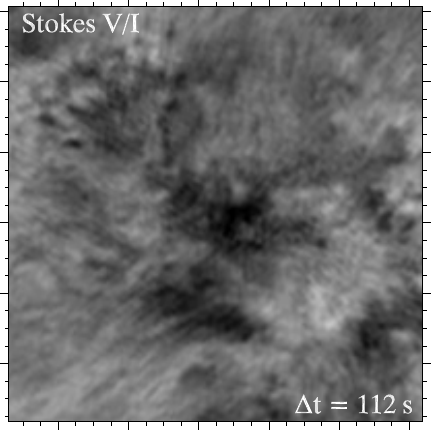}}
       \caption{A time-series showing a subfield of the FOV in
         Stokes~$I$ (the two uppermost rows) and Stokes~$V/I$ (two
         lowermost rows) at $\Delta\lambda=-150$~m\AA. The UF is
         clearly visible in the central part of the series ($\Delta t
         = 48-80$~s). The minor tickmark separation is $0\farcs 5$. An
         online movie is available 
       (\url{http://folk.uio.no/jaime/movies/fig3_delacruzetal2013.mov}).}
        \label{fig:f3}
\end{figure*}

The $\lambda 8542$ data presented here are, to our best knowledge, the
highest-resolution spectroscopic observations of UFs.
Fig.~\ref{fig:f3} shows the detailed evolution of a subfield (marked
in Fig.~\ref{fig:f2} with 
the white square in the lower right corner)
for eight time steps of our
series. In Stokes~$I$ (first and second row), signatures of the UF are
visible from $\Delta t=32$~s to $\Delta t=80$~s, and the sunspot seems
to be in a quiescent state everywhere else. During the UF, our images
reveal dynamic fine structure within the umbra that resemble
penumbral micro-jets \citep{2007katsukawa}, similarly reported by
\citet{SNMC+09}.

In Stokes~$V$, the imprint of the UF is dramatic. During quiescence,
the Stokes~$V$ images show negative polarity everywhere in the
umbra. As the flash starts at $\Delta t=32$, the presence of polarity
changes is tightly correlated with brightenings in Stokes~$I$.
  
Our observations also show the imprint of micro-jets above the penumbra
in Stokes~$V$. They are seen very conspicuously when the time series
is played as a movie.

The Stokes~$Q$ and~$U$ signals are too weak inside the umbra.
However, the fibrils that form the superpenumbra on the limb side of
the spot, show strong signal in Stokes~$Q$ and~$U$ within
$\Delta\lambda=\pm 300$~m\AA. Running penumbral waves are visible in
the four Stokes parameters. In Sect.~\ref{sec:rpw} we analyze
them in more detail.

\subsection{The weak-field approximation for Zeeman-induced
  polarization}\label{sec:weak}
\begin{figure}[]
      \centering
        \resizebox{\hsize}{!}{\includegraphics[trim=0cm 0.0cm 0cm 0.0cm,
          clip]{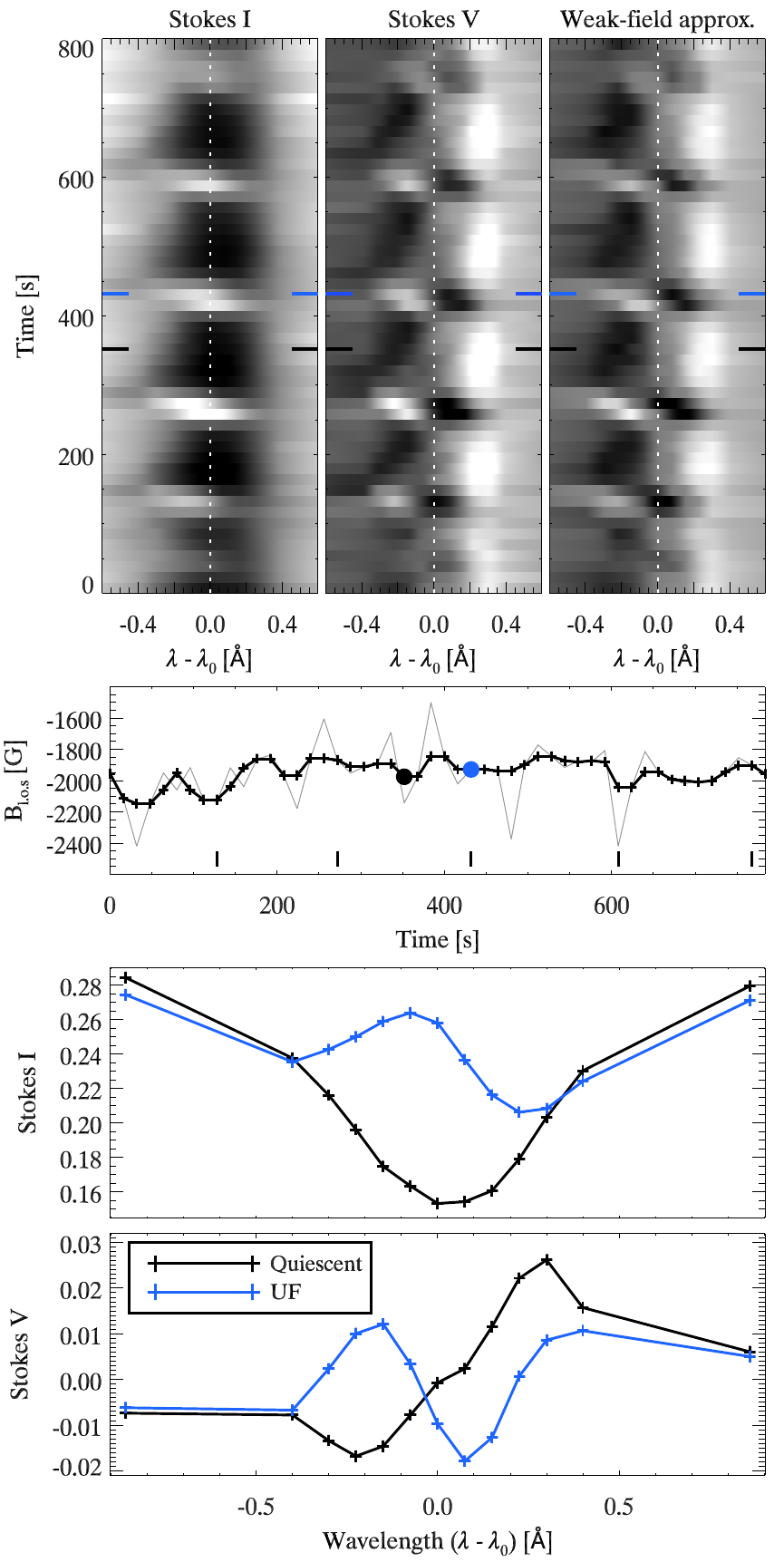}}
       \caption{A time-series at a single pixel within the sunspot is
         shown in the top three-panels, in Stokes~$I$ (top left) and
         $V$ (top middle). The top-right panel corresponds to a
         \emph{synthetic} Stokes~$V$ profile computed using the
         weak-field approximation. The resulting longitudinal magnetic
         field strenght is plotted in the second row. For clarity, the
         lowermost panels illustrate the Stokes~$I$ and~$V$ profiles
         at a quiescent state (black) and during the UF (blue).}
        \label{fig:f4}
\end{figure}

The weak-field approximation provides a relatively simple way of
computing the magnetic field vector from full-Stokes observations
\citep[see][and references therein]{2004egidio}. It works particularly
well in those cases where the Doppler width of the observed line
($\Delta \lambda_D$) is larger than the Zeeman splitting produced by
the magnetic field ($\Delta \lambda_B$). Under these conditions, the
observed Stokes~$Q$, $U$ and~$V$ profiles can be reproduced as a
function of the magnetic field strength and the first derivative of
the Stokes~$I$ profile.

The \ion{Ca}{ii}~$\lambda8542$ line has a relatively low Land\'e
factor ($\bar{g}=1.10$) and it is broader than most photospheric
lines. Therefore, the weak-field approximation can be safely used even
when the magnetic field is relatively strong.  The inequality $\Delta
\lambda_D > \Delta \lambda_B$ provides a convenient means to calculate
the regime of applicability for the approximation
\citep{2011asensio-ramos}:
\begin{equation}
  B < \frac{4\pi m_e c}{\bar{g}\lambda_0 e_0}\sqrt{\frac{2kT}{M} +
    v_\mathrm{mic}^2}\label{eq:bmax}  \, ,
\end{equation}
where $m_e$ and $e_0$ are the electron charge and mass, $\lambda_0$ is the
central wavelength of the line, $c$ is the speed of light, $k$ is the
Boltzmann constant, $M$ is the mass of the atomic element, $T$ is the
temperature and $v_\mathrm{mic}$ the microturbulence.

Assuming a chromospheric scenario with $T=4500$~K,
$v_\mathrm{mic}=3$~km s$^{-1}$ and calcium atoms, the upper limit is
approximately $B\leq 2500$~G. Even though this value is lower than
usual photospheric values in sunspots ($B\approx3000$~G), the magnetic
field strength is usually weaker in the chromosphere.

One limitation of the weak-field approximation is that it also assumes
a constant magnetic field. While most photospheric lines are sensitive
to a relatively thin layer of the atmosphere, where the magnetic field
is nearly constant, the $\lambda8542$ line is sensitive to a range of
approximately 1500~km \citep[e.g.][]{2008cauzzi} and it is less
justified to assume a constant magnetic over that height
range. Another possible approach would be to use the \emph{bisector
  method} \citep{1967rayrole,2004egidio}, which allows one to retrieve
some depth information but it is more sensitive to noise.

In this work we use the weak-field approximation to infer an
approximation to the line-of-sight (l.o.s.) magnetic field strength at
chromospheric heights. To keep a simple approach, we use only $\Delta
\lambda = \pm 225$~m\AA \ from line center.

Fig.~\ref{fig:f4} summarizes our results for one pixel in the sunspot
as a function of time. The top row illustrates $t-\lambda$ diagrams
for Stokes~$I$ and~$V$. The rightmost panel shows the Stokes~$V$
profiles that result from our fits using the weak-field approximation.

The figure shows nicely the Stokes~$V$ polarity reversals during the
UF, which coincide with emission in the Stokes~$I$ profile. In fact, the
shapes of the observed Stokes~$V$ profiles are tightly reproduced by the
first derivative of Stokes~$I$ and the polarity reversals are
consistent with the changes in the slope of Stokes~$I$ that are
induced by the UF.

The top-middle panel in Fig.~\ref{fig:f4} illustrates the inferred
longitudinal magnetic field component during the time series. Despite
some fluctuations with time, the magnetic field strength remains
rather constant showing that it is possible to reproduce UF profiles
with a constant magnetic field. We have indicated all the UF phases
using short vertical lines at the bottom of the panel.

For clarity, we have selected one quiescent profile and one UF
profile, marked in the top row with black and blue horizontal lines,
respectively. Both profiles are plotted in the lowermost two
panels. The corresponding time steps are marked in the upper-middle
panel, showing that the field strength is very similar during the
quiescent and flash phases (at least within the limitations of the
weak-field approximation which provides us with a rough average of the
magnetic field).

These results confirm that the strong variability of the $\lambda8542$
line is likely to be produced by changes in the temperature and
velocity of the plasma and, to a much lesser extent, by changes in the
magnetic field. Note that the field is probably perturbed by the shock
waves during the UF, but it is not a main contributor to that
variability observed in Stokes~$I$ or~$V$.

\section{The inversion code \textsc{Nicole}}\label{sec:inv}

To derive atmospheric parameters from our spectropolarimetric
observations, we use the non-LTE inversion code \textsc{Nicole}
\citep[][Socas-Navarro et al. in
  prep]{2000socas-navarrob}. \textsc{Nicole} solves the non-LTE
problem using preconditioning as explained by
\citet{1997socas-navarro}, assuming that {\em each pixel} harbors a
plane-parallel atmosphere. All Zeeman sub-levels originating from a
given atomic level are assumed to be equally populated, discarding any
quantum interference between them, as proposed by
\citet{1996trujillo-bueno}. We use a \ion{Ca}{ii} model atom
consisting of five bound level plus a continuum
\citep[see][]{2009leenaarts} and the collisional broadening is
computed according to \citet{1998barklem}.

The \emph{velocity-free} approximation is used to compute the atom
population-densities. By neglecting the velocity-field, only half of
the profile needs to be computed and fewer quadrature angles are used
to compute the mean intensity at each depth-point. This approximation
is justified by the very broad shape of the \ion{Ca}{ii} lines and the
fact that the atomic level populations are dominated by the radiation
field at the line core. Once the atomic populations are converged, the
emerging full-Stokes vector is computed including the velocity field
using a quadratic DELO-Bezier formal solution of the polarized
radiative transfer equation \citep{2013delacruz_bezier}.

Scattering polarization is not included in our calculations but this
is not a problem because the magnetic field is sufficiently strong to
produce Zeeman-dominated Stokes profiles. In the $\lambda8542$ line,
it is safe to assume Zeeman induced signal when the magnetic field is
stronger than approximately $100$~G \citep[further details can be
  found in][]{2010manso,2013carlin}.

\textsc{Nicole} has been previously used in LTE studies
\citep{2011socas-navarro,2013scharmer} and validated in non-LTE using
a chromospheric 3D MHD numerical simulation \citep{2012delacruz}.  In
that work, the synthetic profiles are degraded with a realistic CRISP
transmission profile. In order to properly convolve with the
instrumental profile, the synthetic spectra are calculated first in a
fine grid of regularly spaced wavelength points. The inversion code
works internally with such finely sampled profiles except when it
comes to comparing with observations, at which point only the observed
line positions are used to drive the inversion.

The inversion process is initialized with a smoothed version of the
HSRA atmosphere \citep{1971hsra}. Node points are located
equidistantly along the depth-scale, where corrections to the model
are applied. The correction in between nodes is determined as a cubic Bezier-spline interpolation \citep[see][]{2003auer} of the node values.

To improve convergence properties, the inversion is run in two cycles
as proposed by \citet{1992ruiz-cobo}. In the first cycle, fewer
degrees of freedom (nodes) are prescribed. Once the solution cannot be
further improved, a new cycle is started with an (slightly) increased
number of nodes. The degrees of freedom used for each cycle,
summarized in Table~\ref{tab:nodes}, have been obtained after some
trial-and-error experiments with a few sample profiles. Note that we
have used fewer nodes than in \citet{2012delacruz} because our
observations are slightly undersampled around line core, and very few
points are present in the photospheric wings of the line. Therefore,
some depth information is lost with respect to what could be obtained
with more observables.

The equation of state defines a unique relationship among temperature,
electron pressure and gas pressure. If two of them are known, then the
third one can be obtained from it. During the inversion, however, only
the temperature is set for any guess model. Therefore, hydrostatic
equilibrium is used to compute the gas pressure.

A recent study by \citet{2012vannoort} proposes to use the
instrumental point-spread-function (PSF) to couple the inversions
spatially. His method improves the convergence rate of inversions,
producing very smooth results and it allows to resolve details at
the diffraction limit of the telescope. However, it relies on precise
knowledge of the instrumental PSF, which is not always the case in
ground-based observations, where the seeing-related PSF needs to be
taken into account. Here we use the traditional 1.5D approach,
where each pixels in inverted independently from the rest, and
therefore less consistency and more noise is expected in the 2D maps
of inferred quantities.

\begin{table}
  \centering
  \caption{\label{tab:nodes} Summary of node points used during each
    cycle of the inversion. Note that $B_{l.o.s}$ is oriented along
    the $z$-axis.} 
  \begin{tabular}{ r | c c}
    \hline\hline
    Physical parameter & Nodes in cycle 1 & Nodes in cycle 2 \\   
    \hline
    Temperature & 4 & 5\\
    l.o.s velocity & 2 & 3 \\
    Microturbulence & 0 & 0 \\
    $B_{l.o.s} \equiv B_z$          & 0 & 1 \& 2 \\
    $B_x$          & 0 & 1 \\
    $B_y$          & 0 & 1 \\
    \hline
  \end{tabular}
\end{table}

\section{Atmospheric properties during umbral flashes}\label{sec:aum}

To characterize the atmospheric state during UFs, we have inverted two
time steps from the series, corresponding to $\Delta t=16$~s and
$\Delta t=64$~s in Fig.~\ref{fig:f3}. To accelerate the computations,
only every second pixel in both spatial directions was inverted.  We
are not interested in the very small-scale (close to the telescope
diffraction limit) details of the umbra. Therefore, this loss of
resolution (now $0\farcs24$) does not influence our conclusions
significantly whereas computation speed is decreased by a factor of four.

\begin{figure}[]
      \centering
        \resizebox{\hsize}{!}{\includegraphics[trim=0cm 0cm 0cm 0cm,
          clip]{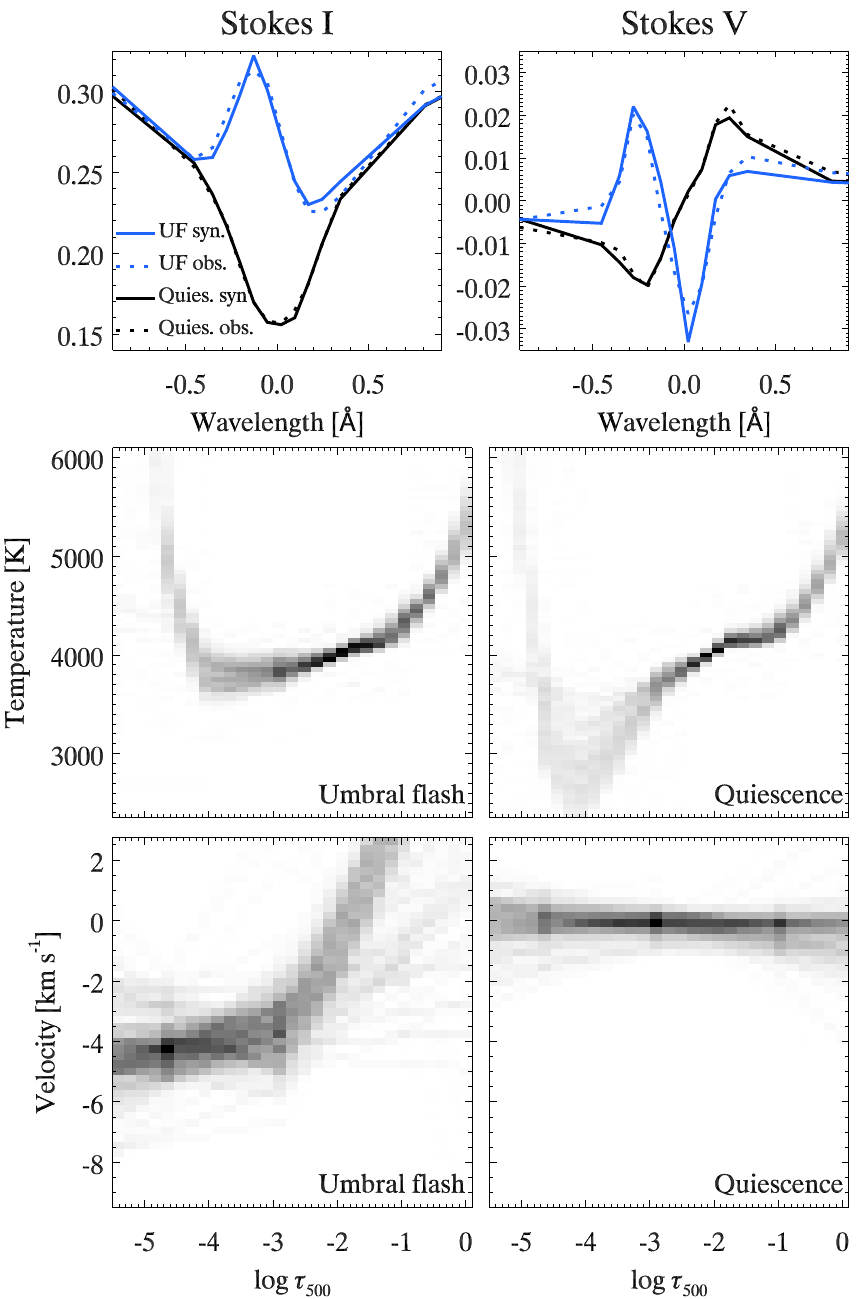}}
       \caption{\emph{Top row:} Observed (dashed line) and inverted (solid
         line) profiles from one pixel
         within the sunspot during the UF phase (blue) and quiescence
         (black). \emph{Middle row:} Density map illustrating the
         inferred temperature stratification during UF (left) and
         quiescence (right) for a patch inside the umbra. A total of 280 models is shown. \emph{Bottom
         row:} Same as middle row, for the l.o.s velocity
       stratification.  
       Negative velocities correspond to upflows.}
        \label{fig:f5}
\end{figure}

\begin{figure}[]
      \centering
          \resizebox{\hsize}{!}{\includegraphics[trim=0cm 0cm 0cm 0cm,
          clip]{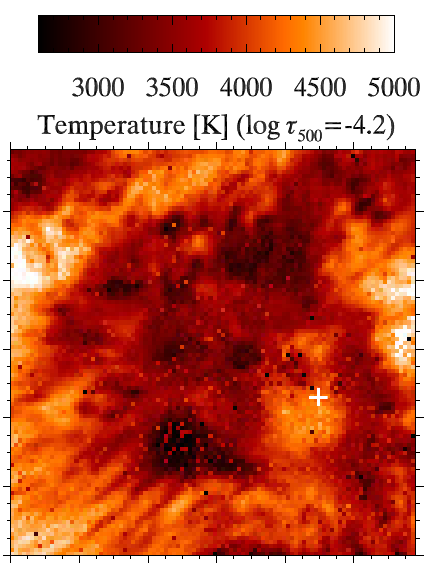}\includegraphics[trim=0cm 0cm 0cm 0cm,
          clip]{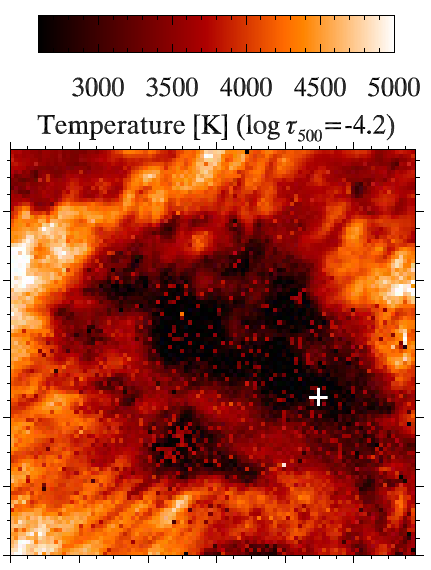}}
          \resizebox{\hsize}{!}{\includegraphics[trim=0cm 0cm 0cm 0cm,
          clip]{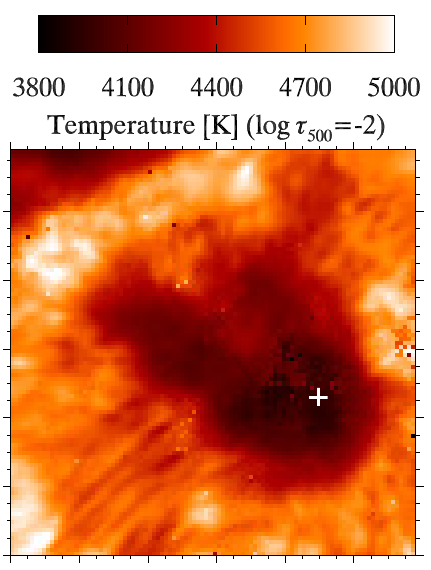}\includegraphics[trim=0cm 0cm 0cm 0cm,
          clip]{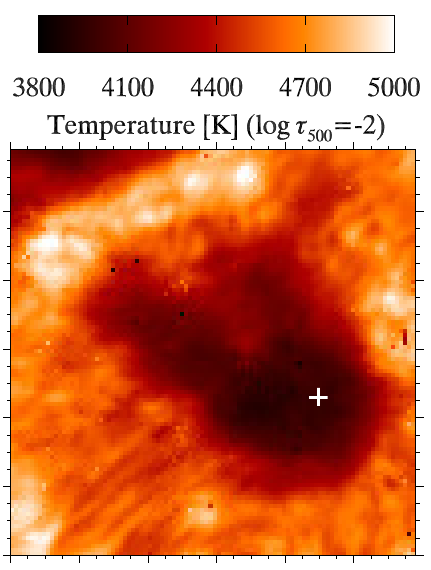}}
          \resizebox{\hsize}{!}{\includegraphics[trim=0cm 0cm 0cm 0cm,
          clip]{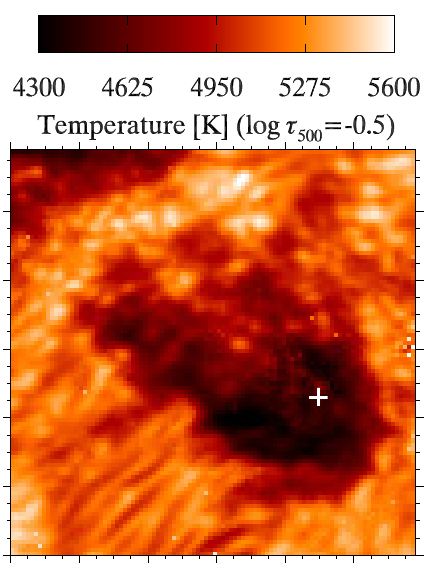}\includegraphics[trim=0cm 0cm 0cm 0cm,
          clip]{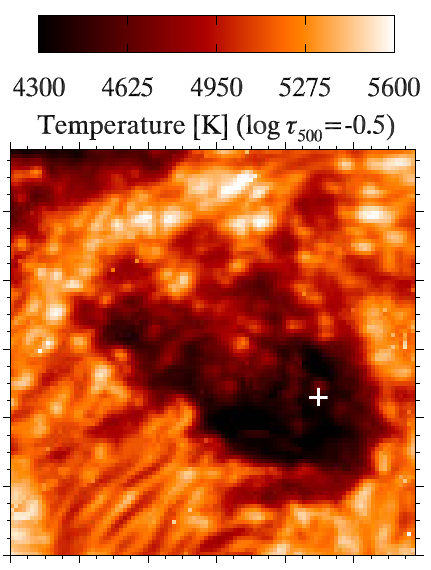}}
       \caption{Inferred temperature from a non-LTE
         inversion. Temperature slices from the inversion are shown in
       two columns at quiescent (right) and flash (left) phases at
       several optical-depth $\log \tau_{500}=-4.2,-2,-0.5$ respectively
       from top to bottom. Note that we have performed a depth average
       of $0.5$~dex, centered at those heights. The minor tick-mark separation is $0\farcs 5$.
       The white cross marks the center of the patch of which the temperature and velocity stratifications are shown in Fig.~\ref{fig:f5}}
        \label{fig:f6}
\end{figure}

Fits to sample profiles are shown in Fig.~\ref{fig:f5}. Observed
(dashed line) and synthetic (solid line) profiles are plotted
for a quiescent phase (black line) and during a UF (blue line) in
Stokes~$I$ (left) and~$V$ (right). Stokes~$Q$ and~$U$ profiles are too
noisy to obtain meaningful fits in the center of the spot. The
inversion code is able to reproduce to a large extent the spectral
features present in the observations, including the emission reversal
seen at line center during the UF. The middle and bottom rows show the inferred quantities as a function
of depth, within a $1\farcs75\times 2\farcs 35$ patch around the
point marked with the cross in Fig.~\ref{fig:f6}, which are discussed below along with 2D
maps of the inferred quantities.

In interpreting inversion results, it is important to understand that
we cannot ``see'' a discrete height. If we take an atmosphere and make
a perturbation (e.g., increase the temperature) in a slab that is much
narrower than the photon mean free path, then the perturbed slab would
be completely transparent and the radiation would travel through it
unchanged. Thus, the spectral profiles emerging from this perturbed
atmosphere would be indistinguishable from the original unperturbed
one. What this means that it is impossible to determine
observationally the properties of a very thin layer. We are sensitive
only to the average properties of the atmosphere on length scales that
are long enough to leave an imprint on the outcoming radiation. This
is not a limitation of the inversion technique or the
instrumentation. It is an intrinsic limitation imposed by the integral
form of the radiative transfer equation. How severe this limitation is
in practice depends on a number of parameters, the most important
being the number of spectral lines observed. Obviously, having several
lines probing the same atmospheric region increases the depth
resolution of the inversions. 

Because of the limitation in height resolution discussed above, it is
not always sensible to interpret inversion results at a specific
height. Instead, one needs to average in the vertical direction over a
slab that is thick enough to have a measurable effect on the observed
profiles. Therefore, the 2D maps presented in this section have been
averaged over a depth range of $0.5$~dex, centered at the $\log\tau_{500}$  value
indicated in each of the panels.

The temperature stratification within the umbra changes dramatically
from the quiescence to the UF phase. The results from our inversions
are illustrated using 2D maps at several optical depths in
Fig.~\ref{fig:f6}. Note that a constant $\tau$ slice usually maps a
corrugated height surface.

In the middle photosphere, at $\log \tau_{500}=-0.5$, there is not a
significant difference between quiescence and UF (lowermost panels of
Fig.~\ref{fig:f6}). We can only see those intensity changes produced
by solar evolution. Some umbral dots are visible within the umbra, and
part of the light bridge is included in the upper-left corner of the
field of view (FOV).

In the upper photosphere (middle panels), the light bridge and
surrounding penumbra become smoother and most of the fine structure of
umbral dots has disappeared.

Signatures of the UF start to appear within the sunspot at
approximately $\log \tau_{500}=-3.0$, where the temperature starts to
increase relative to the quiescent phase. This temperature excess is
very conspicuous at $\log \tau_{500}\approx-4.5$. 
The lower-left panel in Fig.~\ref{fig:f5} shows that the photospheric
temperature is essentially identical during both phases, starting to
diverge approximately at $\log \tau_{500}=-3$. The distributions of
models in this small subfield shows that the
maximum temperature difference between both phases, about 1500~K,
occurs at $\log \tau_{500}=-4$.
 
The 2D temperature maps in the chromosphere (top row in
Fig.~\ref{fig:f6}), show a hot canopy above the umbra during the UF
that is not present during quiescence. This canopy coincides with
those areas showing enhanced line-core emission. 

When shocks are present, we have a very sharp discontinuity of the
physical parameters in the vertical direction. Discontinuities are
blurred out by the radiative transfer and, as discussed above,
observations do not allow to distinguish a sharp change from a
smoother transition. \citep{2012delacruz} demonstrated how in similar
situations, and having only the $\lambda8542$ line, the inversion retrieves a
somewhat smoothed out version of the shock.

\begin{figure}[]
      \centering
          \resizebox{\hsize}{!}{\includegraphics[trim=0cm 0cm 0cm 0cm,
          clip]{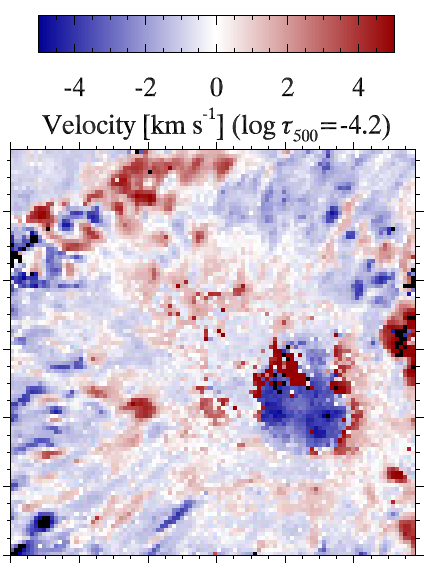}\includegraphics[trim=0cm 0cm 0cm 0cm,
          clip]{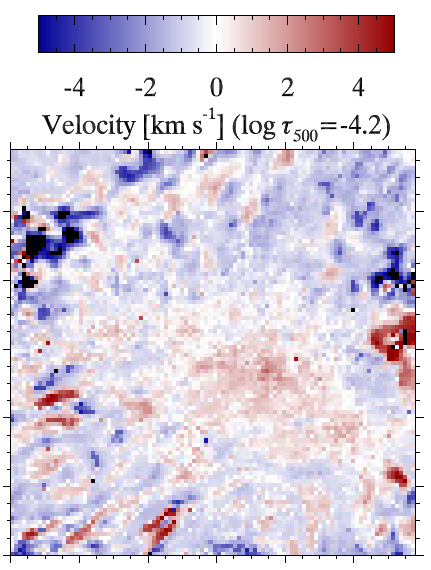}}
          \resizebox{\hsize}{!}{\includegraphics[trim=0cm 0cm 0cm 1.05cm,
          clip]{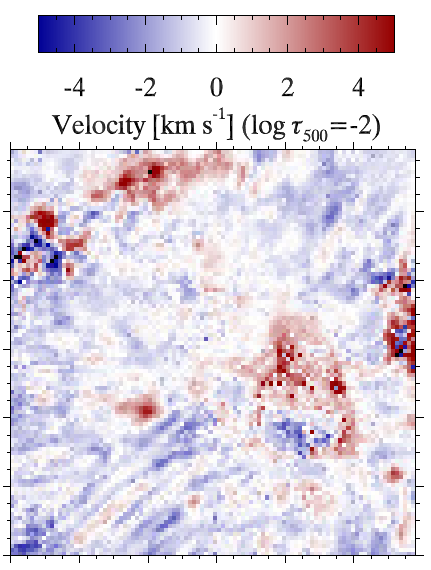}\includegraphics[trim=0cm 0cm 0cm 1.05cm,
          clip]{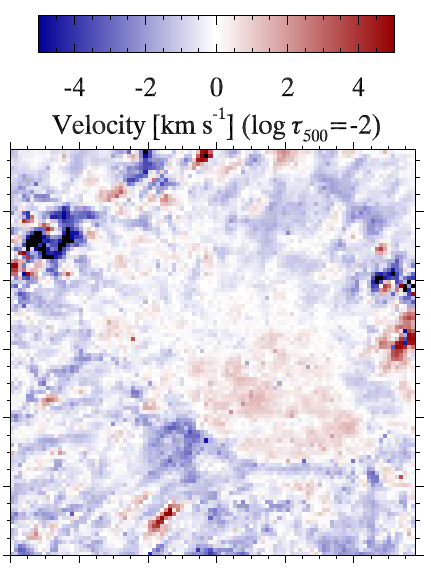}}
       \caption{l.o.s. velocity component from a non-LTE
         inversion. Slices at constant optical depth are shown in 
       two columns at quiescent (right) and flash (left) phases at
       $\log \tau_{500}=-4.2,-2$ respectively
       from top to bottom. Negative velocities correspond to upflows. The minor tick-mark separation is $0\farcs 5$.}
        \label{fig:f7}
\end{figure}

Fig.~\ref{fig:f7} shows the l.o.s velocity component at $\log
\tau_{500}=-4.2$ and $\log \tau_{500}=-2$ from top to bottom
respectively. In the chromosphere (top row), the inversion code
reconstructs upflowing velocities up to 5~km s$^{-1}$ in the flash
(left panel). During the quiescence, the umbra shows an homogeneous
downflow patch ($\sim$1~km s$^{-1}$), similarly to the scenario
reported by \citet{SNTBRC01}.

Another interesting result is found on the light bridge. The time
series shows that micro-jets have apparent up and down motions in the
intensity images. Our maps show that those motions are in
counter-phase with the UF within the umbra, suggesting that the
flash has an effect on the motions of the light bridge, as well.

One limitation of our observing setup is that we have only one
pseudo-continuum point in the far wing of the line. This point allows
us to probe the lower photospheric temperature but, lacking a
symmetric point on the other side of the line, the velocities at that
height are poorly determined. This explains why the temperature map 
is smooth in the photosphere
whereas the corresponding velocity map is much noisier.

The longitudinal component of the magnetic-field reconstructed by the
inversion is noisy. We have compared the results from inversions with
1 node and 2 nodes respectively in $B_{long}$. The resulting
maps are almost identical when a depth average between $\log
\tau_{500}=-4.5$ and $\log \tau_{500}=-3.5$ is considered, although
we notice that the inversion noise increases when the degrees of
freedom is higher.

Further testing with the inversion weights (different for each of the
Stokes parameters) showed that the reduced wavelength coverage in our
observations allows the code to \emph{over-fit} the noisy Stokes~$V$
profiles, which are usually perfectly reproduced by the fitted spectra.

\begin{figure}[]
      \centering
          \resizebox{\hsize}{!}{\includegraphics[trim=0cm 0cm 0cm 0cm,
          clip]{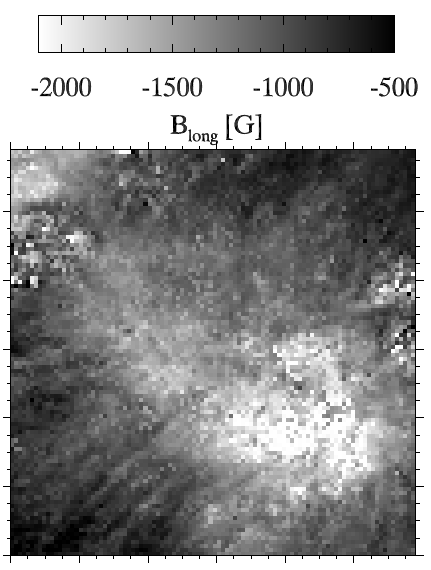}\includegraphics[trim=0cm 0cm 0cm 0cm,
          clip]{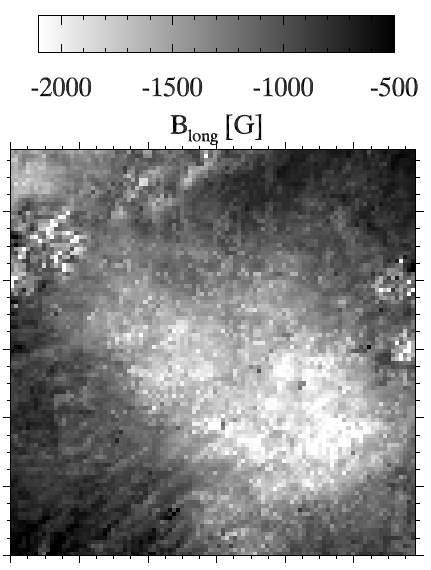}}
       \caption{Longitudinal magnetic field component from a non-LTE
         inversion with one node in $B_{long}$, during the UF
         phase (left) and quiescence (right). The minor tick-mark separation is $0\farcs 5$.}
        \label{fig:f8}
\end{figure}

\begin{figure}[]
      \centering
          \resizebox{\hsize}{!}{\includegraphics[trim=0cm 0cm 0cm 0cm,
          clip]{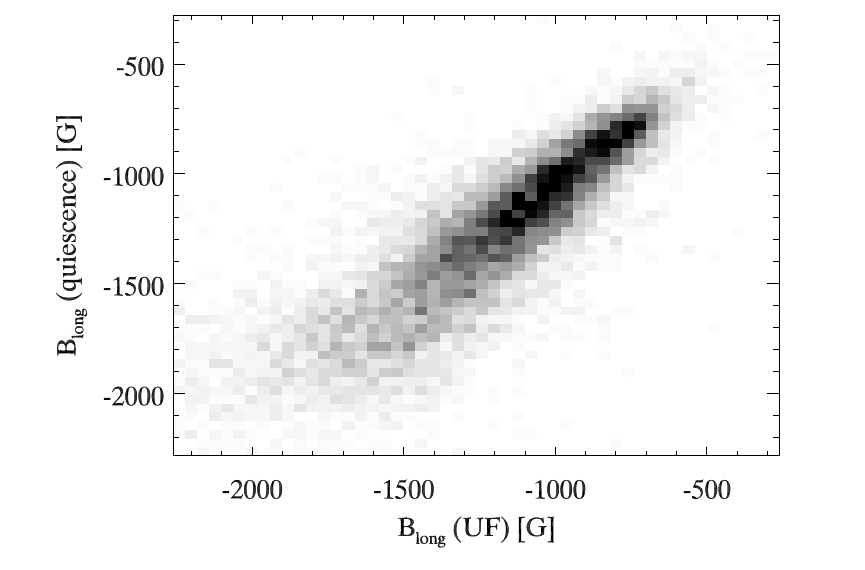}}
       \caption{A 2D probability map illustrating the relation between the
         longitudinal magnetic field component during the UF and
         quiescence, over the inverted field of view.}
        \label{fig:f9}
\end{figure}

Increasing the degrees of freedom makes this situation worse,
therefore we have used the inversion including only 1 node,
illustrated in Fig.~\ref{fig:f8} and Fig.~\ref{fig:f9}. The former
illustrates the resulting $B_{long}$ across the FOV, assuming a constant magnetic
field as a function of depth. As expected, the magnetic field strength
is stronger in the center of the umbra, and becomes weaker
outwards, rather typical configuration for this kind of sunspots. 
Some fibril-like features are visible in the lower-left
corner of the FOV, where the superpenumbra starts. 

Even though we do not find significant changes in the configuration of
the magnetic-field
($B_{long}$) during the flash, small scale variations of approximately 200-300~G are
present within the umbra. Those variations would be compatible with a
noise of $10^{-2}$ in Stokes~$V$ (a rough estimate computed with the
weak-field approximation, the FTS atlas profile
and a magnetic field of 2000~G), however, our observations reach
$10^{-3}$ in the range $\Delta \lambda=\pm300$~m\AA, suggesting that
the source of these variations may not be the noise.

Additionally, small scale features are present over the lightbridge (top-left
corner), indicating a less smooth and more confined magnetic field
configuration, suggesting a connection with the chromospheric jets that
here are also present over the lightbridge.

Fig.~\ref{fig:f9} illustrates a probability diagram comparing $B_{long}
(\mathrm{UF})$ vs. $B_{long} (\mathrm{quiescence})$ over the inverted FOV. This figure
is compatible with an scenario where the magnetic field does not
change much during the UF, as we could somewhat conclude from our
simple analysis using the weak-field approximation.

The lack of signal in Stokes~$Q$ and $U$ inside the umbra does not allow for a
relevant interpretation of the horizontal components of the magnetic
field, however, we include 1 node in $B_x$ and in $B_y$ respectively
so the code can compute a Zeeman splitting compatible with the
observations where the magnetic field strength is close to $2500$~G
(see Sect.~\ref{sec:uf}).

\section{Running penumbral waves}\label{sec:rpw}

\citet{2012delacruz} justify and demonstrate the suitability of LTE
inversions to derive the l.o.s. velocity and the magnetic field vector
from chromospheric observations, disregarding the retrieved
``temperature'' merely as a fitting parameter parameter with no actual
physical meaning. The idea is to allow the inversion code to adjust
the temperature stratification in such a way that the resulting LTE
source function is the same that would be obtained in non-LTE with the
correct temperature. Normally, in \ion{Ca}{ii} inversions this results
in an unrealistically cold atmosphere lacking a transition region. The advantage of this approach
is that one can then analyze longer time-series or larger FOVs as the computing time per inversion is significantly reduced. 

Note
that \citet{2013beck} also perform LTE inversions to retrieve
atmospheric parameters in the upper photosphere using the extended
wings of the \ion{Ca}{ii}~$\lambda3968$~(H line), but those authors
restricted themselves to photospheric quantities.

We have used LTE inversions to study running penumbral waves in our data.
With a similar setup to the one described in Sec.~\ref{sec:inv},
we inverted a small subfield over 193 snapshots. The FOV has been
selected to include a region with strong Stokes~$Q$ and $U$ signal
(thus improving our sensitivity to the transverse field component). The
FOV is indicated in Fig.~\ref{fig:f2} (see the small white box on the
WB image) . This $2\farcs 5 \times 5\arcsec$ patch has been
conveniently selected such that the running penumbral waves propagate
parallel to the longest dimension.

\begin{figure}[]
      \centering
          \resizebox{\hsize}{!}{\includegraphics[trim=0cm 0cm 0cm 0cm,
          clip]{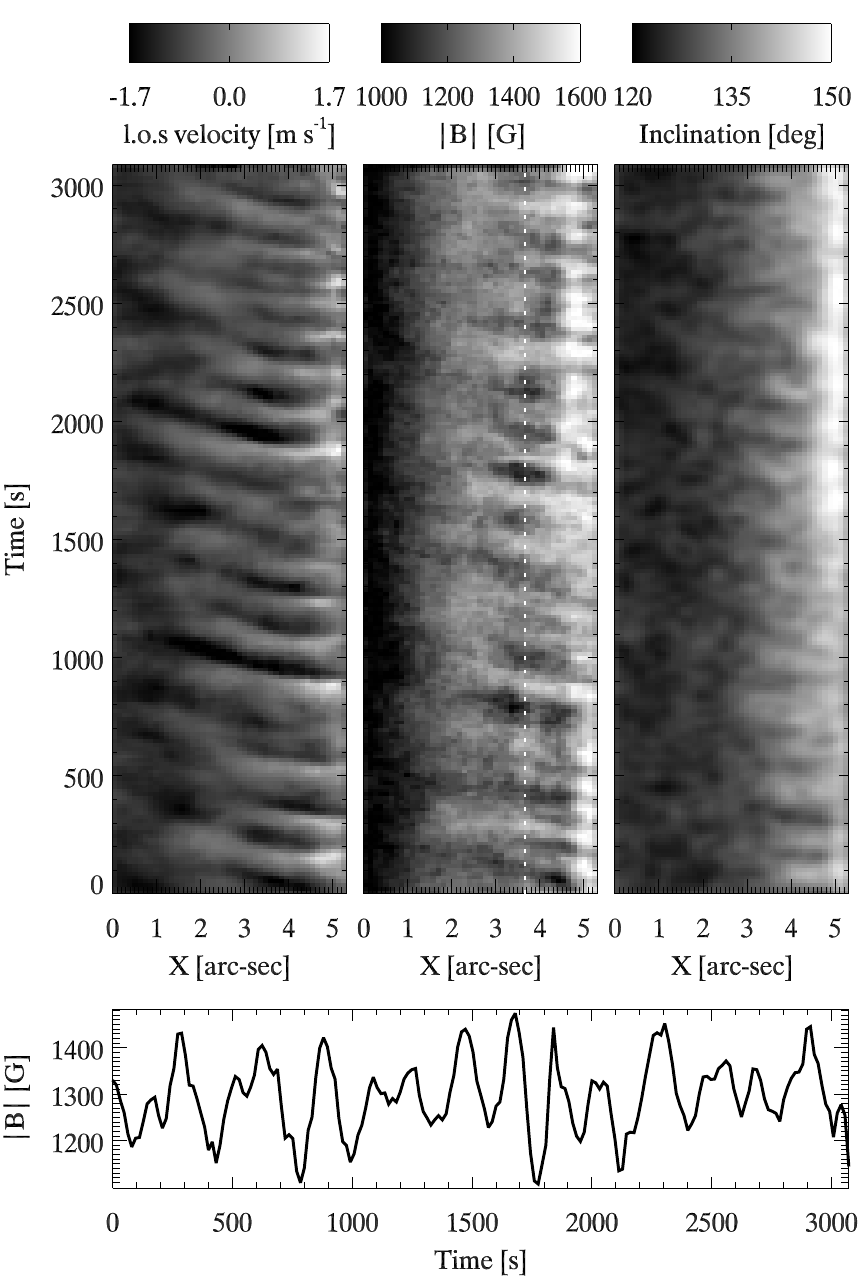}}
       \caption{$x$-$t$ diagrams for the l.o.s velocity
         (left), the magnetic field strength (middle) and the magnetic
       field inclination (right) are shown in the top row. The umbra is on the right side of the panels, the outer penumbra on the left side ($x=0$).
       The detailed magnetic field strength modulation for one spatial location (at $x=3.65$, indicated with the white dotted line in the middle panel) is plotted in the bottom panel.
       The
       magnetic field inclination has been smoothed with a 3 pixels
       full-width-half-maximum Gaussian kernel.}
        \label{fig:f10}
\end{figure}

\begin{figure}[]
      \centering
          \resizebox{\hsize}{!}{\includegraphics[trim=0cm 0cm 0cm 0cm,
          clip]{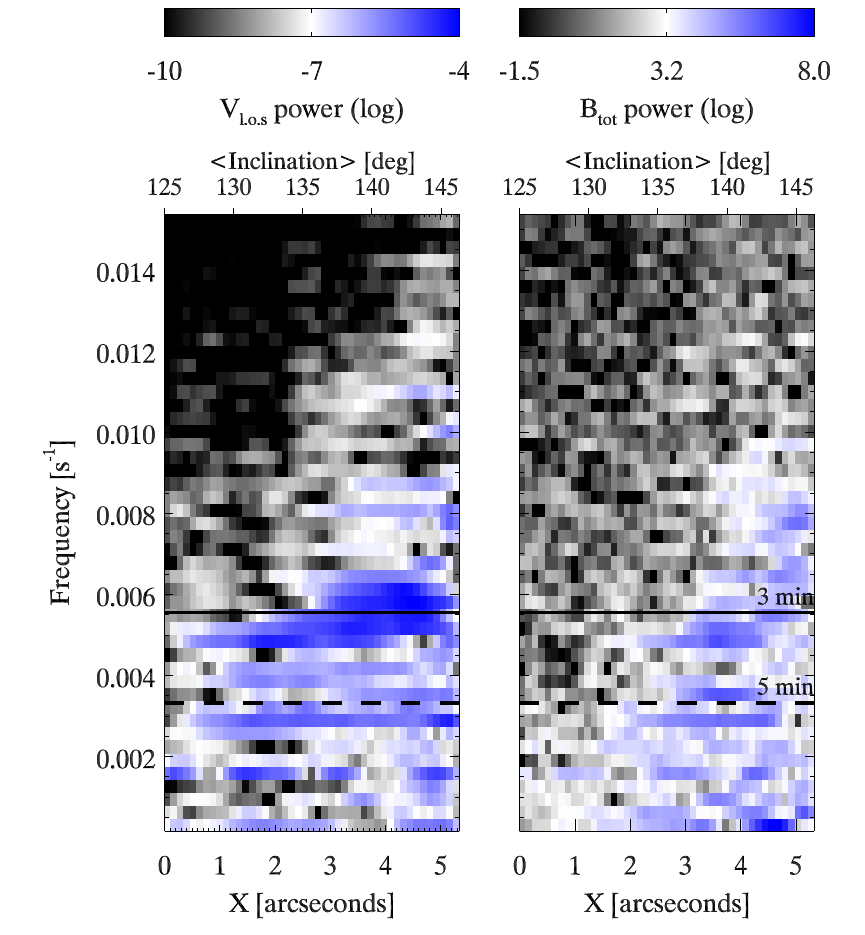}}
       \caption{FFT power spectra along each column for the l.o.s
         velocities and magnetic field strength presented in
         Fig.~\ref{fig:f10}. The axis labeling at the top indicates the average magnetic field inclination obtained through averaging over the right panel of Fig.~\ref{fig:f10}. More inclined field is found in the outer penumbra ($x=0$), more vertical field in the umbra (right side of the panels). The solid horizontal line indicates the 3 minutes mode, whereas the horizontal dashed line indicates the 5 minutes mode.}
        \label{fig:f11}
\end{figure}

The resulting magnetic field vector has been converted into the solar
reference frame to compute the local inclination of the field
\citep[similarly done in][]{2013scharmer}. We
solve the $180\deg$ ambiguity by selecting, at each pixel, the
direction that makes the field point radially outwards in the local
frame. 
In Fig.~\ref{fig:f10} and \ref{fig:f11} 
we average the plotted quantities in
the direction perpendicular to the wave propagation (the small
dimension of the box) with the aim of increasing the signal-to-noise
ratio.  The l.o.s. velocity and longitudinal magnetic field are
depth-dependent. For those quantities we plot depth averages from $\log
\tau_{500}=-3.5$ to $\log \tau_{500}=-4.5$.

The top panels of Fig.~\ref{fig:f10} 
show the l.o.s velocity (left),
magnetic field strength (middle) and the magnetic field inclination
(right) as a function of time. 
Waves are visible in these diagrams as a modulation pattern of alternating bright and dark features. In the penumbra (left part of the panels), these features are slanted which is indicative of wave propagation.
These are the running penumbral waves which are so prominently visible in the Stokes $I$ and $V$ movies.
In the umbra ($x>5$) the modulation pattern is more horizontal and is related to UFs.
The slanted features are clearly visible in the l.o.s. velocity and, with somewhat lower
contrast, in the longitudinal field as well.
The slope of the features indicate a propagation speed of 
about 13~km~s$^{-1}$,
which is higher but consistent with the measurements of 
\citet{RvdVRS+03} 
who analyzed running penumbral waves in \ion{Ca}{ii}~H and K images.

According to the results shown in the middle panel, the magnetic field
strength is perturbed by the running penumbral wave, showing
peak-to-peak variations up to 200~G. For clarity, the time series
along the white dotted line is shown at the bottom panel.

The rightmost panel in Fig.~\ref{fig:f10} 
 illustrates the evolution of
the magnetic field inclination as a function of time. The magnetic
field is almost vertical close to the umbra and it rapidly becomes
more horizontal outwards. Some modulation may be discerned as a
function of time, but it is very weak and the fore-mentioned diagonal
slope in the pattern is not clear. We then conclude that the field
inclination is not perturbed significantly within our detection
threshold (a few degrees) as the wave propagates. This result suggests
that the shock front has an effect on the field strength but not on its
orientation.

The FFT power spectra along each of the columns of the l.o.s velocities
and magnetic field strength (see Fig.~\ref{fig:f10}) are presented in
Fig.~\ref{fig:f11}. 
We find that in both cases, higher frequencies are more dominant towards the umbra (right side of the panels)
where the magnetic field is more vertical (as indicated by the average magnetic field inclination derived from averaging over columns of the right panel of Fig.~\ref{fig:f10} and marked on the top axis).
Outwards into the penumbra (lower values of $x$), most power is concentrated in
lower frequency modes, producing a slope in the power spectra diagram.

\section{Conclusions}\label{sec:disc}
In this paper we present a novel analysis of state-of-the-art
spectropolarimetric observations of a sunspot chromosphere. In
particular, we have focused on the study of the physical parameters of
the chromosphere during UFs.

A simple analysis of the Stokes~$V$ reversals that are observed during
UFs show that, even with a nearly constant magnetic field, the
profiles can be successfully reproduced. Despite the limitations of
the weak-field approximation, it is possible to connect the variability of
Stokes~$V$ with the changes of the intensity profile (Stokes~$I$).

The resulting model from our non-LTE inversions suggest a temperature
excess up to 1000~K in the shocked chromosphere, relative to the quiescence
phase. In the photosphere, the temperature stratification seems
unaffected by the UF phenomenon.

To reproduce the observed profiles, a steep velocity gradient is
reconstructed at $\log \tau_{500} \approx -3$. Above, the velocity remains
almost constant around $v_{l.o.s} = 4.5$~km s$^{-1}$. These velocities
are similar to those from the simulation by \citet{2010bard} at
approximately 1000~km height.

In the
photosphere, the line is very broad and our observations have limited coverage of
photospheric wavelengths. Therefore, the inversion is rather
insensitive to large photospheric velocities, which rather are
adjusted to reproduce the fore-mentioned velocity gradient.

Our observations do not allow for a detailed interpretation of the horizontal
component of the magnetic field inside the umbra:
the umbral Stokes $Q$ and $U$ signals are too weak to allow for robust results. 
Therefore we only
comment on the longitudinal component. The overall magnetic field
configuration remains rather constant inside the umbra during the
flash, but we find subarcsecond fluctuations that could be produced
by the flashes.

The situation is slightly different in the penumbra, though, perhaps
because the field is more inclined 
(resulting in a stronger component perpendicular to the l.o.s.)
or because it is weaker (or a
combination of both reasons). We inverted a 2D subfield along the
superpenumbra where running penumbral waves were clearly observed. The
magnetic field vector inferred by the inversions is almost vertical at
the boundary of the umbra, becoming more horizontal outwards. Further
analysis of the entire time series have allowed us to measure the
imprint of these waves in the magnetic field strength, which seems to
be modulated with a peak-to-peak amplitude some 200~G. Our results,
however, do not show a similar fluctuation of the magnetic field
inclination, which suggests that the wavefront is able to change the
field strength but not its geometry.

We find a trend of decreasing high-frequency modulation power for more inclined magnetic fields for the l.o.s. velocity and magnetic field strength. In the outer penumbra we find an absence of high frequency power while there is increasingly more power at high frequencies towards the umbra. 
This observation is interesting in the context of wave propagation along inclined magnetic fields. 
Low frequency waves ($p$-modes), that are normally evanescent in the low density chromosphere, can leak into the chromosphere along inclined magnetic fields 
\citep{1990LNP...367..211S, 
2004Natur.430..536D}. 
This mechanism has been identified as the driving mechanism of dynamic fibrils
\citep{2006ApJ...647L..73H, 
2007ApJ...655..624D, 
2007ApJ...666.1277H}. 
This mechanism might well be at play for UFs and running penumbral waves too and our analysis demonstrates that the kind of observations we present here are well suited to study this in more detail. 
We leave the extension of the inversions to a wider FOV and inclusion of a wider range of inclination angles that are needed to be more conclusive on this matter to future work.

\begin{acknowledgements}
The Swedish 1-m Solar Telescope was operated by the Institute for Solar
Physics of the Royal Swedish Academy of Sciences in the Spanish
Observatorio del Roque de los Muchachos of the Instituto de
Astrof\'{\i}sica de Canarias. We thank Eamon Scullion for valuable help
during the observations. Part of the computations were performed on
resources at Chalmers Centre for Computational Science and Engineering
(C3SE) provided by the Swedish National Infrastructure for Computing
(SNIC), with project number SNIC002-12-27.
 \end{acknowledgements}

\appendix
\section{The flat-field correction}\label{sec:descatter}
Especial care must be taken with the
flat-fielding procedure. At the SST, the flat-field data are acquired
while the telescope moves in circles around solar disk center,
accumulating $\approx 1000$ images per wavelength and polarization
state. Effectively, the data represent a temporal and
spatial average of quiet-Sun data at solar disk center. Therefore, it
is reasonable to assume that the observed line profile at each pixel
should be almost identical, except for the dirt, wavelength dependent
fringes and the prefilter transmission, that can change across the FOV.

The telecentric configuration used in CRISP
is optimal to achieve high spatial-resolution observations, although
the instrumental transmission profile is not constant across the
field-of-view (FOV). Pixel-to-pixel wavelength shifts (cavity errors) are present due
to imperfections in the surface of the etalons that cannot be
infinitely flat. Additionally, imperfections in the coating of the
etalons produce reflectivity variations across the FOV that change
the width of the transmission profile (reflectivity errors).

These effects become obvious when a spectral line is present at the
observed wavelength, producing intensity fluctuations across the FOV 
that correlate tightly with the cavity errors and, less obviously, with
variations in the reflectivity.  

While these instrumental effects are 
permanently anchored to the same pixels on the camera, solar
features are moved around by differential seeing motions, however, the MOMFBD
code is only able to process data where the intensity at any given
pixel, is strictly changing due to atmospheric distortion. To overcome this
problem \citet{2011schnerr} proposed a strategy to
characterize CRISP and prepare flat-field images that allow to reconstruct the
data. 

Their idea is founded on the assumption that each pixel observed \emph{the
same quiet-Sun profile} during the flat-field data acquisition. To
remove the imprint of the quiet-Sun profile from each pixel, an
iterative self-consistent scheme is used to separate the different
contributions to the intensity profile at each pixel, which we
summarize in the following lines. To reproduce the observed profile on
a pixel-to-pixel basis, they assume that the intensity at each pixel
can be expressed as,
\begin{equation}
  I_{obs} (\lambda) = c_0\cdot (I_{QS}(\lambda)\ast T(e_{ch},e_{rh})) \cdot(1+c_3*\lambda),
\end{equation}
where $I_{obs}$ is the observed profile at a given pixel, $I_{QS}$ is
the estimate of the spatially-averaged quiet-Sun profile, $T$ is a theoretical 
CRISP transmission profile, $c_0$ is a multiplicative factor that
scales the entire profile and $c_3$ accounts for shifts of the
prefilter. The $\ast$ stands for a convolution operator.

We simplify their strategy, by assuming constant reflectivities across
the FOV. Neglecting the effect of reflectivity errors works
particularly well in broad-lines like the $\lambda8542$, given that
the transmission profile is relatively narrow compared to the Doppler
width of the line. It also reduces the time required to compute the
fits, from hours to minutes on a single CPU, as no convolutions are
required any longer. We also allowed to include two more terms to the
prefilter polynomial, given that at this wavelength there are fringes
that change as a function of wavelength, introducing some distortion
in the profiles.

\begin{figure}[]
      \centering
        \resizebox{\hsize}{!}{\includegraphics[trim=0cm 0.0cm 0cm 0.0cm,
          clip]{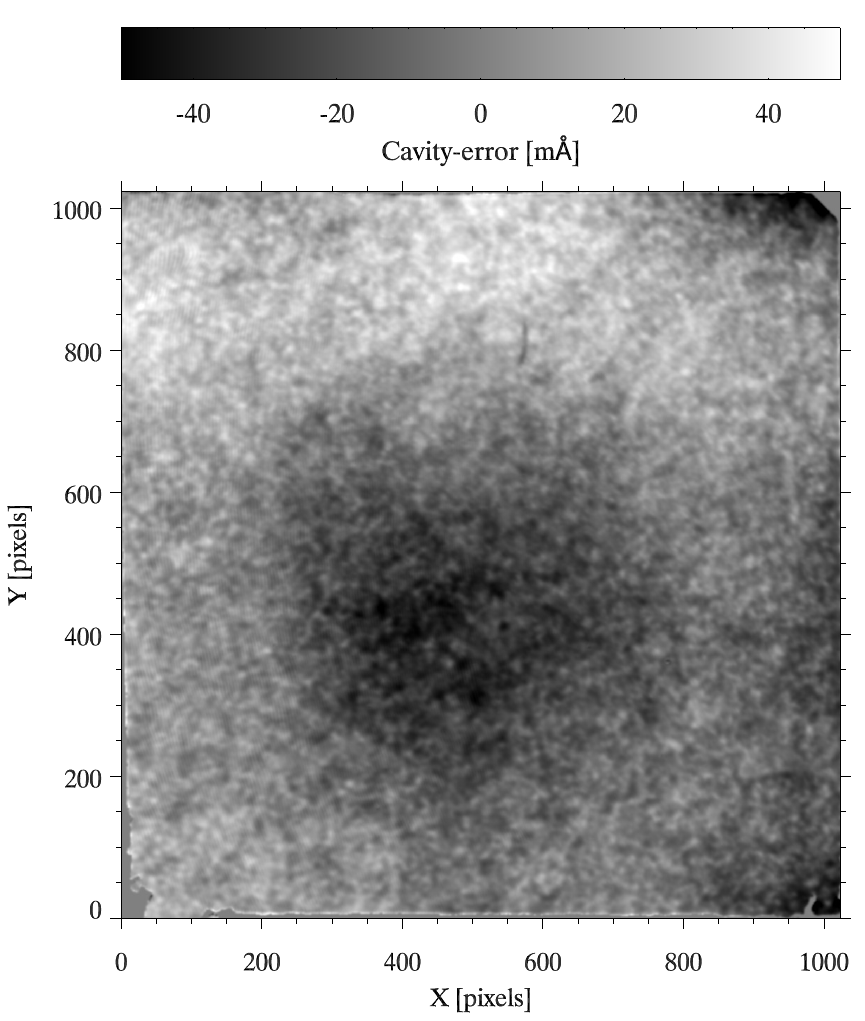}}
        \resizebox{\hsize}{!}{\includegraphics[trim=0cm 0.0cm 0cm 0.0cm,
          clip]{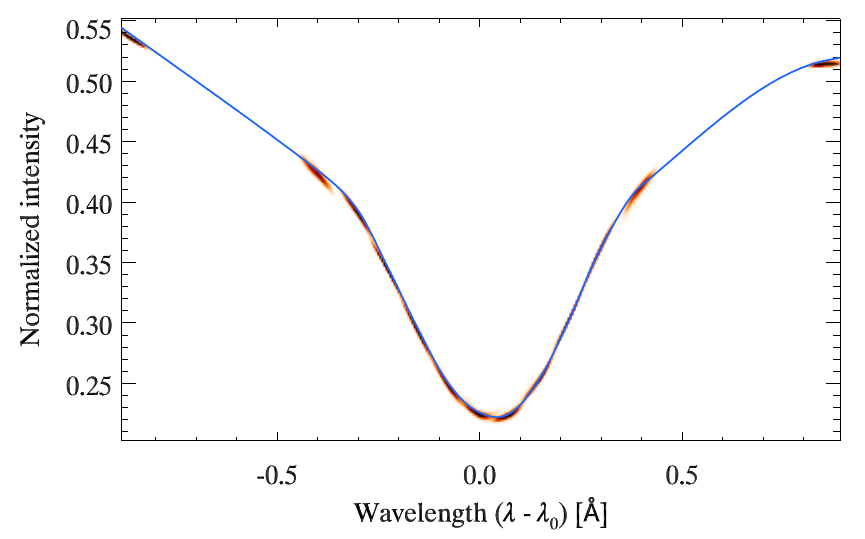}}
       \caption{\emph{Top:}  Map of the FPI cavity errors in units of wavelength shift.
       \emph{Bottom:} Density plot of all pixels in the summed flat-field data (orange clouds) that have been shifted to the correct wavelength and corrected for the different contributions in the model. The blue line is the average line profile described by a spline curve and closely resembles a disk center quiet Sun atlas profile.} 
        \label{fig:ap1}
\end{figure}

Fig.~\ref{fig:ap1} (lower panel) illustrates a density map, where the profiles from
all pixels have been divided by
$c_0\cdot(1+c_3\lambda+c_4\lambda^2+c_5\lambda^3)$ and shifted to the
correct wavelength. The best estimate of the quiet-sun profile is
plotted with a blue line. A tight correlation means that our model is
able to reproduce the observed profiles, despite the simplifications
applied to the original model by \citet{2011schnerr}. A cavity-error
map is also shown in Fig.~\ref{fig:ap1}

\subsection{The CCD backscatter problem}

The back-illuminated 
Sarnoff CAM1M100 
CCD cameras used in the CRISP instrument allow to
acquire 35 frames-per-second, freezing atmospheric motions during each
acquisition. These cameras work particularly well in visible light,
but their efficiency decreases steeply in the infrared. 

\begin{figure}[]
      \centering
        \resizebox{\hsize}{!}{\includegraphics[trim=0cm 0.0cm 0cm 0.0cm,
          clip]{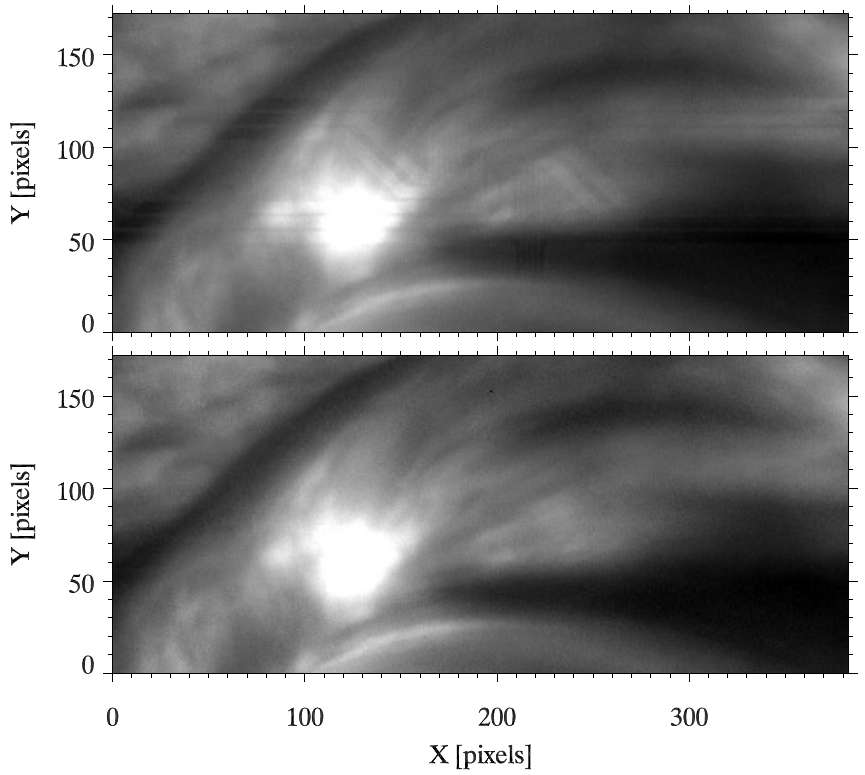}}

       \caption{\emph{Top:} subfield from an individual acquisition, where traditional
         dark-current and flat-field corrections have been applied. 
         	An electronic circuit pattern from the 
         CCD chip is visible in the background. 
         \emph{Bottom:}
         same subfield, backscatter corrected.}
        \label{fig:ap2}
\end{figure}

Above approximately $7000$~\AA,
the CCD becomes semi-transparent and the images show a circuit pattern
that is not present at other wavelengths. \citet{2010delacruz}
explains a way to overcome this issue and properly apply the
flat-field correction to the data. Furthermore, the images show a
circuit-like pattern that cannot be removed by traditional dark-field
and flat-field corrections, as illustrated in Fig. \ref{fig:ap2} (top panel).\\
Examination of pinhole data and flat-fielded data shows that:
\begin{enumerate}
	\item There is a diffuse additive stray-light contribution in the whole image.
	\item The stray-light contribution is much smaller in the circuit pattern and the gain appears to be enhanced.
\end{enumerate}

\begin{figure}[]
      \centering
      \resizebox{\hsize}{!}{\includegraphics[]{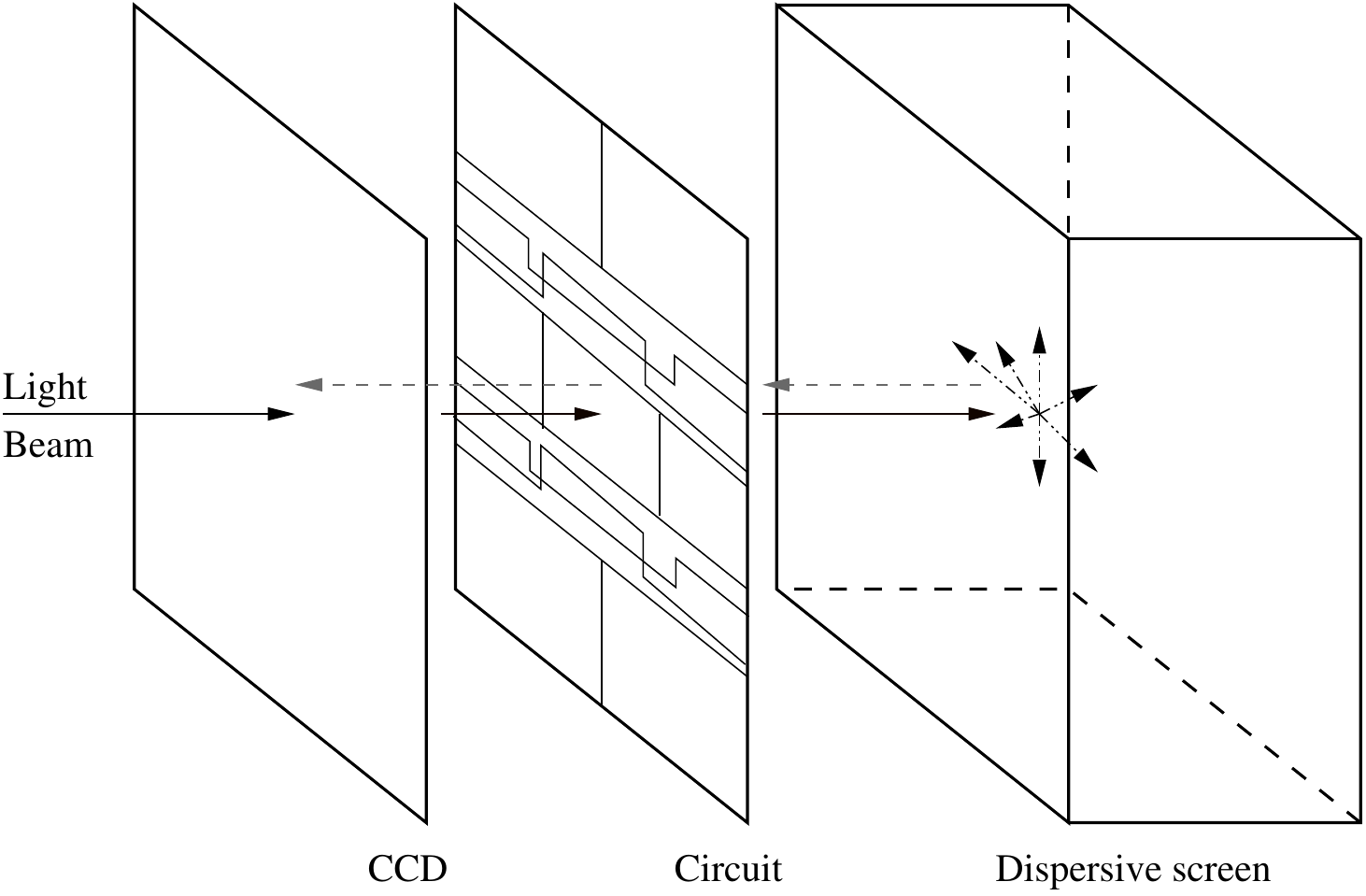}}
        \caption{
		sketch showing a conceptual model of the camera. The dark arrows indicate the incoming light beam from the telescope and the gray arrows represent the back-scattered light that returns to the CCD. 
       }
        \label{fig:ap3}
\end{figure}

These problems can be explained by a semi-transparent CCD with a diffusive medium behind it, combined with an electronic circuit located right behind the CCD that is partially reflecting and therefore also less transparent to the scattered light.  Under normal conditions an image recorded with a CCD, $I_o$, can be described in terms of the dark-field ($D$), the gain factor ($G_f$) and the real image ($I_r$):
\begin{equation}
	I_o=D+(G_f \cdot I_r).\label{nimg}
\end{equation}
However, in the infrared we need a more complicated model:
\begin{equation}
	I_o = D +f(1 - f )[(G_b G_f I_r ) * P] G_b + f G_f I_r \label{sfull}
\end{equation}
where $f$ represents the overall fraction of light absorbed by the
CCD, $G_b$ is the gain for light illuminating the CCD from the back
which should account for the electronic circuit pattern. In the
following, we refer to $G_b$ as \emph{backgain}. $P$ is a Point Spread
Function (PSF) that describes the scattering properties of the
dispersive screen. In the backscatter term we assume that
$(1-f)G_fI_rG_b$ is transmitted to the diffusive medium where it is
scattered. A fraction of the scattered light returns to the CCD,
passing again through the circuit. The sketch in  Fig.~\ref{fig:ap3} shows a schematic representation of the structure of the camera and the path followed by the light beam.

\begin{figure}[]
      \centering
      \resizebox{\hsize}{!}{\includegraphics[]{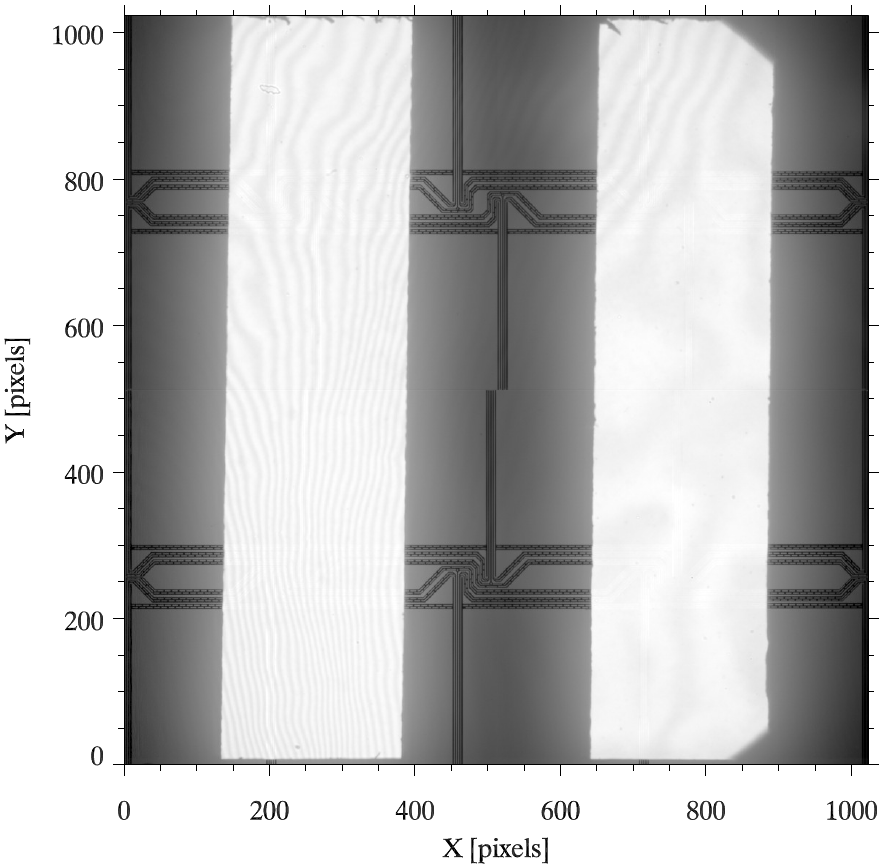}}
        \caption{
	Calibration data acquired to infer the backgain and the PSF
        describing the backscatter problem. The dark areas are
        physically masked on the focal plane of the telescope. The
        color scale is saturated to enhance the CCD chip pattern.
       }
        \label{fig:ap4}
\end{figure}

To obtain the real intensity $I_r$, the PSF $P$, the back-gain $G_b$ and the front-gain $G_f$ must be known. The transparency factor is assumed to be smooth because the properties of the dispersive screen seem to be homogeneous across the field-of-view. This allows us to include $f$ in the front and the back gain factors
$$G_{b^\prime}=\sqrt{1-f} \ G_b$$
$$G_{f^\prime}=f G_f,$$
so Eq. \ref{sfull} becomes:
\begin{equation}
	I_o=D+G_{b^\prime}[(G_{b^\prime} G_{f^\prime} I_r)* P] + G_{f^\prime} I_r\label{sfullsim}
\end{equation}
This problem is linear and invertible, but a direct inversion would be expensive, given the dimensions of the problem. A numerical approach can be used to iteratively solve the problem. We define
\begin{equation}
\hat{J} = (G_{b^\prime} G_{f^\prime} I_r)* P
\end{equation}
In the first iteration, we initialize $\hat{J}$ assuming that the smearing caused by $P$ is so large that $\hat{I}$ can be approximated by the \textit{average} value of the observed image multiplied with the back gain. The back gain $G_{b^\prime}$ is assumed to be 1 for every pixel in the first iteration. Furthermore, we assume that the product $G_{f^\prime}I_r$ can be estimated from Eq. \ref{nimg} ignoring back scattering, i.e., $G_{f^\prime}I_r \approx I_o -D$. These values are of the same order of magnitude as the final solution, and therefore correspond to  a reasonable choice of initialization.
\begin{equation}
\hat{J}_0 \approx \langle G_{b^\prime} G_{f^\prime} I_r \rangle \approx \langle G_{b^\prime} (I_o - D) \rangle,
\end{equation}
where the initial guess of the PSF $P$ represents an angular average obtained from a pinhole image. The small diameter of the pinhole only allows to estimate accurately the central part of the PSF. The wings of our initial guess are extrapolated using a power-law. 
The estimate of $G_{f^\prime} I_r$ is then
\begin{equation}
G_{f^\prime} I_r = I_o -D -G_{b^\prime} \hat{J}
\end{equation}
which can be used to compute a new estimate $\hat{J}$. This procedure is iterated, until $G_{f^\prime} I_r$ and $I_o$ are consistent.
This new value of $\hat{J}$ is used to improve our estimate of $G_{b^\prime}$, by applying the same procedure to images that contain parts  being physically masked ($I_r = 0$), as in Fig. \ref{fig:ap4}. 
In the masked parts where $I_r\equiv 0$ we have,
\begin{equation}
I_o - D - G_{b^\prime} \hat{J} = 0,
\end{equation}
so the back-gain can be computed directly:
\begin{equation}
G_{b^\prime}  = \frac{I_o-D}{\hat{J}}.\label{backg}
\end{equation}
Thus, every pixel must have been covered by the mask at least once in a calibration image in order to allow the calculation of the back-gain. With the new $G_{b^\prime}$ we can recompute a new estimate of $G_{f^\prime} I_r$.

We now need to specify a measure that describes how well the data are fitted by the estimate of $P$ and $G_{b^\prime}(P)$. Since both $G_{f^\prime} I_r$ and $G_{b^\prime}$  are computed based on self-consistency, we iteratively need to fit only the parameters of the PSF. We assume that the PSF is circular-symmetric and apply corrections to the PSF at \emph{node} points placed along the radius.

We use Brent's Method described by \citet{2002nr} to minimize our fitness function. This algorithm does not require the computation of derivatives with respect to the free parameters of the problem. When the opaque bars block a region of the CCD, an estimate of the back gain can be calculated for a given PSF according to Eq.~\ref{backg}. In our calibration data, the four bars of width $L$ are displaced $0.5 \ L$ from one image to the next. This overlapping provides two different measurements of the back gain on each region of the CCD. However, as the location of the bars changes on each image, the scattered light contribution is different for each of these measurements of the back gain. Our fitness function minimizes the difference between these two measurements of the back gain. 

Having thus obtained the PSF $P$ and the back-gain $G_{b^\prime}$, we obtain $G_{f^\prime}$ by recording conventional flats and assuming $I_r$ is a constant in order to obtain $G_{f^\prime}$ from Eq.~\ref{sfullsim}.


\section{The SST polarization model at $\lambda854$~nm}\label{sec:telescope}
The optical path of the SST, from the 1m lens down to the cameras,
contains optical elements that can polarize light. Therefore,
polarimetric observations at the SST must be calibrated.

At the SST this calibration is split in two parts: the optical table
and the turret. The former is calibrated daily, given that the
observer can easily place a linear polarizer and a retarder to
generate known polarization states, and analyze how the optical table
changes the polarization \citep[see][for further details]{2008vannoort}. 

The result is a modulation
matrix $\mathbf{M}_{table}$ that
allows to convert the observed polarization states $\mathbi{I}_{lc}\equiv(I_{lc0}, I_{lc1},
I_{lc2}, I_{lc3})^T$ into Stokes parameters $\mathbi{I}\equiv(I,Q,U,V)^T$
parameters, using a linear combination:
\begin{equation}
\mathbi{I}'=\mathbf{M}^{-1}_{table}\cdot \mathbi{I}_{lc},
\end{equation}
where the prime symbol indicates that only the optical table has been
calibrated. 

In principle, if the telescope was built with non-polarizing elements
then $(I',Q',U',V')\equiv(I,Q,U,V)$, but this is not the
case. The turret has mirrors and a lens that change the polarization properties
of the incoming light. In fact, the polarizing properties of the
turret change with the pointing. Therefore, a final transformation
($\mathbf{M}_{tel}$) is
needed to obtain telescope-polarization free Stokes parameters:
\begin{equation}
\mathbi{I}=\mathbf{M}^{-1}_{tel}\cdot \mathbi{I}'\equiv \mathbf{M}^{-1}_{tel}
\cdot (\mathbf{M}^{-1}_{table}\cdot \mathbi{I}_{lc})
\end{equation}

\citet{2010selbing} studied the
polarizing properties of the SST and proposed a model to characterize
the telescope induced polarization at 630~nm. In this work, we use a
similar model. Their work allows to compute $\mathbf{M}_{tel}$ based
on the azimuth and elevation parameters of the telescope pointing.

Calibration images have been recorded during an entire day, using a 1-m polarizer mounted at
the entrance lens of the SST. The polarizer rotates $360\deg$ in
steps of $5\deg$ and several frames are acquired for each angle. These
data are used to derive the parameters of the model proposed by
\citet{2010selbing} at 854~nm. This model is already presented in
\citet{2010delacruz} and included here for completeness.

Each polarizing optical element of the telescope is represented with a
Mueller matrix. $\mathbf{L}$ represents the entrance lens. The form of
the lens matrix represents a composite of random retarders, so it
demodulates the light without polarizing it. The reference for $Q$ is
aligned with the 1m linear polarizer axis and the values of the matrix
are measured in that frame:
$$
\mathbf{L}(l_0,l_1,l_2,l_3,l_4) =  {\footnotesize\left( \begin{array}{c c c c}
		1		&	0		&	0		&	0 \\
		0		&	l_0		&	l_1		&-l_2 \\
		0		&	l_1		&	l_3		&l_4\\
		0		& 	l_2		& -l_4	&	(l_0+l_3-1)
	\end{array} \right)}.
$$

Mirrors are noted with $\mathbf{M}$ and have two free
parameters. Assuming a wave that propagates along the $z$-axis and
oscillates in the $x-y$ plane, the parameters are the de-attenuation
term $R$ between the $xy$ components of the electromagnetic wave and
the phase retardance $\delta$ produced by the mirror. This form of
$\mathbf{M}$ assumes that $Q$ is perpendicular to the plane of incidence:
$$
	\mathbf{M}(R,\delta)  =  {\footnotesize\left( \begin{array}{c c c c}
		\frac{1}{2}+\frac{1}{2}R	&		\frac{1}{2}-\frac{1}{2}R	&	0		&	0 \\
		\frac{1}{2}-\frac{1}{2}R	&		\frac{1}{2}+\frac{1}{2}R	&	0		&	0 \\
		0		& 0	&	-\sqrt{R}\cos (\frac{\pi}{180}\delta)	&	-\sqrt{R}\sin(\frac{\pi}{180}\delta)	 \\
		0		& 0	&	\sqrt{R}\sin (\frac{\pi}{180}\delta)	&	-\sqrt{R}\cos(\frac{\pi}{180}\delta)	 \\
	\end{array}\right)}.
$$

Finally, $\mathbf{R}(\alpha)$ corresponds to a rotation to a new
coordinate frame:
$$
	\mathbf{R}(\alpha) = {\footnotesize \left( \begin{array}{c c c c}
		1		&	0		&	0		&	0 \\
		0		&	\cos(\frac{\pi}{90}\alpha) &	\sin (\frac{\pi}{90}\alpha)	&	0 \\
		0		&	-\sin(\frac{\pi}{90}\alpha)	&	\cos (\frac{\pi}{90}\alpha)	&	0 \\
		0		& 0	&	0	&	1
	\end{array} \right)}
$$
\citet{2010selbing} builds the telescope model using the Mueller
matrix of each polarizing element, as a
function of the azimuth $(\varphi)$ and elevation $(\theta)$ angles of
the Sun at the time of the observation (Eq.~\ref{eq:telmod}). We have included the
conversion factor from degrees to radians in the matrices, thus all
the angles are given in degrees:
\begin{equation}
\begin{split}
	M_{tel}\left(\theta,\varphi\right) = & R_{f+}(c_{11}) \cdot
        M_{s}(c_9,c_{10})\cdot 
        M_{f}(c_7,c_8) \cdot\\
        &\cdot R_{f-}(c_{11})\cdot R_{az}(\varphi)
        \cdot M_{az}(c_5,c_6) \cdot
       R_{el}(\theta)\\ 
        & \cdot M_{el}(c_5,c_6) \cdot
        L(c_0,c_1,c_2,c_3,c_4), \label{eq:telmod}
\end{split}
\end{equation}
where $c_{1,2,...,11}$ are the parameters of the model, computed using
a least-squares fit to the observed data. The fitted values are
summarized in Table~\ref{tab:fitval}.
 \begin{table}
 \centering
 \caption{Parameters of the telescope model obtained from calibration
   data at 8542~nm.}
 \begin{tabular}{c | l}
   \hline
   \hline
 	Parameter & Fitted value \\
 	\hline 
	$c_0$  &		$+9.596161\cdot 10^{-1}$		\\
	$c_1$  &		$-2.541791\cdot 10^{-3}$			\\
	$c_2$  &		$-8.337375\cdot 10^{-3}$	\\
	$c_3$  &		$+9.640901\cdot 10^{-1}$		\\
	$c_4$  &		$-1.875988\cdot 10^{-2}$		\\
	$c_5$  &		$+9.164327 \cdot 10^{-1}$		\\
	$c_6$  & 		 $+1.465364\cdot 10^1$\\
	$c_7$  &		$+1.017471\cdot 10^0$		\\
	$c_8$  &		$+1.326977\cdot 10^0$	\\
	$c_9$  &		 $+1.010035\cdot 10^0$	\\
	$c_{10}$ &		 $-1.592178\cdot 10^0$	\\
	$c_{11}$ &		 $+4.000000\cdot 10^0$	\\
        \hline
\end{tabular}
\label{tab:fitval}
\end{table}

Note that the polarizer has a significant leak of unpolarized light at 854.2 nm. Using small samples of the sheets used to construct the 1-m polarizer, we have estimated the extinction ratio of the 1-m polarizer to be approximately $40\%$, so the parameters of the lens have been determined assuming that value. 

 The linear polarization reference is defined by the first mirror after the lens, however this is not very useful in practice because the turret introduces image rotation along the day. Instead, we use the solar north as a reference for positive $Q$ by applying an extra rotation to $\mathbf{M}_{tel}$.

The full-Stokes \ion{Ca}{ii} images in Fig.~\ref{fig:f2} have been
calibrated using the telescope model described in this section.

\bibliographystyle{aa}
\bibliography{spot.bib,hsn.bib}

\begin{thebibliography}{61}
\expandafter\ifx\csname natexlab\endcsname\relax\def\natexlab#1{#1}\fi

\bibitem[{{Asensio Ramos}(2011)}]{2011asensio-ramos}
{Asensio Ramos}, A. 2011, \apj, 731, 27

\bibitem[{{Auer}(2003)}]{2003auer}
{Auer}, L. 2003, in Astronomical Society of the Pacific Conference Series, Vol.
  288, Stellar Atmosphere Modeling, ed. {I.~Hubeny, D.~Mihalas, \& K.~Werner},
  3

\bibitem[{{Bard} \& {Carlsson}(2010)}]{2010bard}
{Bard}, S. \& {Carlsson}, M. 2010, \apj, 722, 888

\bibitem[{{Barklem} \& {O'Mara}(1998)}]{1998barklem}
{Barklem}, P.~S. \& {O'Mara}, B.~J. 1998, \mnras, 300, 863

\bibitem[{{Beck} {et~al.}(2013){Beck}, {Rezaei}, \& {Puschmann}}]{2013beck}
{Beck}, C., {Rezaei}, R., \& {Puschmann}, K.~G. 2013, ArXiv e-prints: 1302.6936

\bibitem[{{Beckers} \& {Tallant}(1969)}]{BT69}
{Beckers}, J.~M. \& {Tallant}, P.~E. 1969, \solphys, 7, 351

\bibitem[{{Bharti} {et~al.}(2013){Bharti}, {Hirzberger}, \& {Solanki}}]{BHS13}
{Bharti}, L., {Hirzberger}, J., \& {Solanki}, S.~K. 2013, \aap, 552, L1

\bibitem[{{Bloomfield} {et~al.}(2007){Bloomfield}, {Lagg}, \&
  {Solanki}}]{BLS07}
{Bloomfield}, D.~S., {Lagg}, A., \& {Solanki}, S.~K. 2007, \apj, 671, 1005

\bibitem[{{Brault} \& {Neckel}(1987)}]{fts-atlas}
{Brault}, J.~W. \& {Neckel}, H. 1987, Spectral Atlas of Solar Absolute
  Disk-averaged and Disk-Center Intensity from 3290 to 12510 \AA,
  ftp://ftp.hs.uni-hamburg.de/pub/outgolng/FTS-Atlas

\bibitem[{{Bray} \& {Loughhead}(1974)}]{1974bray}
{Bray}, R.~J. \& {Loughhead}, R.~E. 1974, {The solar chromosphere} (The
  International Astrophysics Series, London: Chapman and Hall, 1974)

\bibitem[{{Carlin} {et~al.}(2013){Carlin}, {Asensio Ramos}, \& {Trujillo
  Bueno}}]{2013carlin}
{Carlin}, E.~S., {Asensio Ramos}, A., \& {Trujillo Bueno}, J. 2013, \apj, 764,
  40

\bibitem[{{Cauzzi} {et~al.}(2008){Cauzzi}, {Reardon}, {Uitenbroek},
  {Cavallini}, {Falchi}, {Falciani}, {Janssen}, {Rimmele}, {Vecchio}, \&
  {W{\"o}ger}}]{2008cauzzi}
{Cauzzi}, G., {Reardon}, K.~P., {Uitenbroek}, H., {et~al.} 2008, \aap, 480, 515

\bibitem[{{Centeno} {et~al.}(2005){Centeno}, {Socas-Navarro}, {Collados}, \&
  {Trujillo Bueno}}]{CSNc+05}
{Centeno}, R., {Socas-Navarro}, H., {Collados}, M., \& {Trujillo Bueno}, J.
  2005, \apj, 635, 670

\bibitem[{{Collados} {et~al.}(1999){Collados}, {Rodr{\'{\i}}guez Hidalgo},
  {Bellot Rubio}, {Ruiz Cobo}, \& {Soltau}}]{CRHBR+99}
{Collados}, M., {Rodr{\'{\i}}guez Hidalgo}, I., {Bellot Rubio}, L., {Ruiz
  Cobo}, B., \& {Soltau}, D. 1999, in Astronomische Gesellschaft Meeting
  Abstracts, 13

\bibitem[{{de la Cruz Rodriguez}(2010)}]{2010delacruz}
{de la Cruz Rodriguez}, J. 2010, PhD thesis, Stockholm University,
  \url{http://urn.kb.se/resolve?urn=urn:nbn:se:su:diva-43646}

\bibitem[{{de la Cruz Rodr{\'{\i}}guez} {et~al.}(2011){de la Cruz
  Rodr{\'{\i}}guez}, {Kiselman}, \& {Carlsson}}]{2011delacruz}
{de la Cruz Rodr{\'{\i}}guez}, J., {Kiselman}, D., \& {Carlsson}, M. 2011,
  \aap, 528, A113

\bibitem[{{de la Cruz Rodr{\'{\i}}guez} \&
  {Piskunov}(2013)}]{2013delacruz_bezier}
{de la Cruz Rodr{\'{\i}}guez}, J. \& {Piskunov}, N. 2013, \apj, 764, 33

\bibitem[{{de la Cruz Rodr{\'{\i}}guez} {et~al.}(2012){de la Cruz
  Rodr{\'{\i}}guez}, {Socas-Navarro}, {Carlsson}, \&
  {Leenaarts}}]{2012delacruz}
{de la Cruz Rodr{\'{\i}}guez}, J., {Socas-Navarro}, H., {Carlsson}, M., \&
  {Leenaarts}, J. 2012, \aap, 543, A34

\bibitem[{{De Pontieu} {et~al.}(2004){De Pontieu}, {Erd{\'e}lyi}, \&
  {James}}]{2004Natur.430..536D}
{De Pontieu}, B., {Erd{\'e}lyi}, R., \& {James}, S.~P. 2004, \nat, 430, 536

\bibitem[{{De Pontieu} {et~al.}(2007){De Pontieu}, {Hansteen}, {Rouppe van der
  Voort}, {van Noort}, \& {Carlsson}}]{2007ApJ...655..624D}
{De Pontieu}, B., {Hansteen}, V.~H., {Rouppe van der Voort}, L., {van Noort},
  M., \& {Carlsson}, M. 2007, \apj, 655, 624

\bibitem[{{Felipe} {et~al.}(2010){Felipe}, {Khomenko}, \&
  {Collados}}]{2010felipe}
{Felipe}, T., {Khomenko}, E., \& {Collados}, M. 2010, \apj, 719, 357

\bibitem[{{Felipe} {et~al.}(2011){Felipe}, {Khomenko}, \&
  {Collados}}]{2011felipe}
{Felipe}, T., {Khomenko}, E., \& {Collados}, M. 2011, \apj, 735, 65

\bibitem[{{Gingerich} {et~al.}(1971){Gingerich}, {Noyes}, {Kalkofen}, \&
  {Cuny}}]{1971hsra}
{Gingerich}, O., {Noyes}, R.~W., {Kalkofen}, W., \& {Cuny}, Y. 1971, \solphys,
  18, 347

\bibitem[{{Hansteen} {et~al.}(2006){Hansteen}, {De Pontieu}, {Rouppe van der
  Voort}, {van Noort}, \& {Carlsson}}]{2006ApJ...647L..73H}
{Hansteen}, V.~H., {De Pontieu}, B., {Rouppe van der Voort}, L., {van Noort},
  M., \& {Carlsson}, M. 2006, \apjl, 647, L73

\bibitem[{{Heggland} {et~al.}(2007){Heggland}, {De Pontieu}, \&
  {Hansteen}}]{2007ApJ...666.1277H}
{Heggland}, L., {De Pontieu}, B., \& {Hansteen}, V.~H. 2007, \apj, 666, 1277

\bibitem[{{Henriques}(2012)}]{2012henriques}
{Henriques}, V.~M.~J. 2012, \aap, 548, A114

\bibitem[{Henriques(2013)}]{thesis-vasco}
Henriques, V. M.~J. 2013, PhD thesis, Stockholm University,
  \url{http://urn.kb.se/resolve?urn=urn:nbn:se:su:diva-86798}

\bibitem[{{Katsukawa} {et~al.}(2007){Katsukawa}, {Berger}, {Ichimoto}, {Lites},
  {Nagata}, {Shimizu}, {Shine}, {Suematsu}, {Tarbell}, {Title}, \&
  {Tsuneta}}]{2007katsukawa}
{Katsukawa}, Y., {Berger}, T.~E., {Ichimoto}, K., {et~al.} 2007, Science, 318,
  1594

\bibitem[{{Kneer} {et~al.}(1981){Kneer}, {Mattig}, \& {Uexkuell}}]{KMU81}
{Kneer}, F., {Mattig}, W., \& {Uexkuell}, M.~V. 1981, \aap, 102, 147

\bibitem[{{Kosugi} {et~al.}(2007){Kosugi}, {Matsuzaki}, {Sakao}, {Shimizu},
  {Sone}, {Tachikawa}, {Hashimoto}, {Minesugi}, {Ohnishi}, {Yamada}, {Tsuneta},
  {Hara}, {Ichimoto}, {Suematsu}, {Shimojo}, {Watanabe}, {Shimada}, {Davis},
  {Hill}, {Owens}, {Title}, {Culhane}, {Harra}, {Doschek}, \&
  {Golub}}]{Hinode1}
{Kosugi}, T., {Matsuzaki}, K., {Sakao}, T., {et~al.} 2007, \solphys, 243, 3

\bibitem[{{L{\' o}pez Ariste} {et~al.}(2001){L{\' o}pez Ariste},
  {Socas-Navarro}, \& {Molodij}}]{LASNM01}
{L{\' o}pez Ariste}, A., {Socas-Navarro}, H., \& {Molodij}, G. 2001, \apj, 552,
  871

\bibitem[{{Landi Degl'Innocenti} \& {Landolfi}(2004)}]{2004egidio}
{Landi Degl'Innocenti}, E. \& {Landolfi}, M., eds. 2004, Astrophysics and Space
  Science Library, Vol. 307, {Polarization in Spectral Lines}

\bibitem[{{Leenaarts} {et~al.}(2009){Leenaarts}, {Carlsson}, {Hansteen}, \&
  {Rouppe van der Voort}}]{2009leenaarts}
{Leenaarts}, J., {Carlsson}, M., {Hansteen}, V., \& {Rouppe van der Voort}, L.
  2009, \apjl, 694, L128

\bibitem[{{Manso Sainz} \& {Trujillo Bueno}(2010)}]{2010manso}
{Manso Sainz}, R. \& {Trujillo Bueno}, J. 2010, \apj, 722, 1416

\bibitem[{Press {et~al.}(2002)Press, Teukolsky, Vetterling, \&
  Flannery}]{2002nr}
Press, W.~H., Teukolsky, S.~A., Vetterling, W.~T., \& Flannery, B.~P. 2002,
  {Numerical recipes in C++: The art of scientific computing}, 2nd edn.
  (Cambridge University Press)

\bibitem[{{Rayrole}(1967)}]{1967rayrole}
{Rayrole}, J. 1967, Annales d'Astrophysique, 30, 257

\bibitem[{{Rouppe van der Voort} {et~al.}(2003){Rouppe van der Voort},
  {Rutten}, {S{\"u}tterlin}, {Sloover}, \& {Krijger}}]{RvdVRS+03}
{Rouppe van der Voort}, L.~H.~M., {Rutten}, R.~J., {S{\"u}tterlin}, P.,
  {Sloover}, P.~J., \& {Krijger}, J.~M. 2003, \aap, 403, 277

\bibitem[{{Ruiz Cobo} \& {del Toro Iniesta}(1992)}]{1992ruiz-cobo}
{Ruiz Cobo}, B. \& {del Toro Iniesta}, J.~C. 1992, \apj, 398, 375

\bibitem[{{Scharmer}(2006)}]{2006scharmer}
{Scharmer}, G.~B. 2006, \aap, 447, 1111

\bibitem[{{Scharmer} {et~al.}(2003){Scharmer}, {Bjelksjo}, {Korhonen},
  {Lindberg}, \& {Petterson}}]{2003scharmer}
{Scharmer}, G.~B., {Bjelksjo}, K., {Korhonen}, T.~K., {Lindberg}, B., \&
  {Petterson}, B. 2003, in Society of Photo-Optical Instrumentation Engineers
  (SPIE) Conference Series, Vol. 4853, Society of Photo-Optical Instrumentation
  Engineers (SPIE) Conference Series, ed. S.~L. {Keil} \& S.~V. {Avakyan},
  341--350

\bibitem[{{Scharmer} {et~al.}(2012){Scharmer}, {de la Cruz Rodr{\'{\i}}guez},
  {S{\"u}tterlin}, \& {Henriques}}]{2013scharmer}
{Scharmer}, G.~B., {de la Cruz Rodr{\'{\i}}guez}, J., {S{\"u}tterlin}, P., \&
  {Henriques}, V.~M.~J. 2012, ArXiv e-prints: 1211.5776

\bibitem[{{Scharmer} {et~al.}(2008){Scharmer}, {Narayan}, {Hillberg}, {de la
  Cruz Rodr{\'{\i}}guez}, {L{\"o}fdahl}, {Kiselman}, {S{\"u}tterlin}, {van
  Noort}, \& {Lagg}}]{2008scharmer}
{Scharmer}, G.~B., {Narayan}, G., {Hillberg}, T., {et~al.} 2008, \apjl, 689,
  L69

\bibitem[{{Schnerr} {et~al.}(2011){Schnerr}, {de La Cruz Rodr{\'{\i}}guez}, \&
  {van Noort}}]{2011schnerr}
{Schnerr}, R.~S., {de La Cruz Rodr{\'{\i}}guez}, J., \& {van Noort}, M. 2011,
  \aap, 534, A45

\bibitem[{{Selbing}(2010)}]{2010selbing}
{Selbing}, J. 2010, ArXiv e-prints: 1010.4142

\bibitem[{{Shine} {et~al.}(1994){Shine}, {Title}, {Tarbell}, {Smith}, {Frank},
  \& {Scharmer}}]{1994shine}
{Shine}, R.~A., {Title}, A.~M., {Tarbell}, T.~D., {et~al.} 1994, \apj, 430, 413

\bibitem[{{Socas-Navarro}(2011)}]{2011socas-navarro}
{Socas-Navarro}, H. 2011, \aap, 529, A37

\bibitem[{{Socas-Navarro} {et~al.}(2009){Socas-Navarro}, {McIntosh}, {Centeno},
  {de Wijn}, \& {Lites}}]{SNMC+09}
{Socas-Navarro}, H., {McIntosh}, S.~W., {Centeno}, R., {de Wijn}, A.~G., \&
  {Lites}, B.~W. 2009, \apj, 696, 1683

\bibitem[{{Socas-Navarro} \& {Trujillo Bueno}(1997)}]{1997socas-navarro}
{Socas-Navarro}, H. \& {Trujillo Bueno}, J. 1997, \apj, 490, 383

\bibitem[{{Socas-Navarro} {et~al.}(2000{\natexlab{a}}){Socas-Navarro},
  {Trujillo Bueno}, \& {Ruiz Cobo}}]{SNTBRC00c}
{Socas-Navarro}, H., {Trujillo Bueno}, J., \& {Ruiz Cobo}, B.
  2000{\natexlab{a}}, \apj, 544, 1141

\bibitem[{{Socas-Navarro} {et~al.}(2000{\natexlab{b}}){Socas-Navarro},
  {Trujillo Bueno}, \& {Ruiz Cobo}}]{SNTBRC00b}
{Socas-Navarro}, H., {Trujillo Bueno}, J., \& {Ruiz Cobo}, B.
  2000{\natexlab{b}}, Science, 288, 1396

\bibitem[{{Socas-Navarro} {et~al.}(2000{\natexlab{c}}){Socas-Navarro},
  {Trujillo Bueno}, \& {Ruiz Cobo}}]{2000socas-navarrob}
{Socas-Navarro}, H., {Trujillo Bueno}, J., \& {Ruiz Cobo}, B.
  2000{\natexlab{c}}, \apj, 530, 977

\bibitem[{{Socas-Navarro} {et~al.}(2001){Socas-Navarro}, {Trujillo Bueno}, \&
  {Ruiz Cobo}}]{SNTBRC01}
{Socas-Navarro}, H., {Trujillo Bueno}, J., \& {Ruiz Cobo}, B. 2001, \apj, 550,
  1102

\bibitem[{{Suematsu}(1990)}]{1990LNP...367..211S}
{Suematsu}, Y. 1990, in Lecture Notes in Physics, Berlin Springer Verlag, Vol.
  367, Progress of Seismology of the Sun and Stars, ed. Y.~{Osaki} \&
  H.~{Shibahashi}, 211

\bibitem[{{Thomas}(1984)}]{T84}
{Thomas}, J.~H. 1984, \aap, 135, 188

\bibitem[{{Trujillo Bueno} \& {Landi
  Degl'Innocenti}(1996)}]{1996trujillo-bueno}
{Trujillo Bueno}, J. \& {Landi Degl'Innocenti}, E. 1996, \solphys, 164, 135

\bibitem[{{Tsiropoula} {et~al.}(2000){Tsiropoula}, {Alissandrakis}, \&
  {Mein}}]{TAM00}
{Tsiropoula}, G., {Alissandrakis}, C.~E., \& {Mein}, P. 2000, \aap, 355, 375

\bibitem[{{Tsuneta} {et~al.}(2008){Tsuneta}, {Ichimoto}, {Katsukawa}, {Nagata},
  {Otsubo}, {Shimizu}, {Suematsu}, {Nakagiri}, {Noguchi}, {Tarbell}, {Title},
  {Shine}, {Rosenberg}, {Hoffmann}, {Jurcevich}, {Kushner}, {Levay}, {Lites},
  {Elmore}, {Matsushita}, {Kawaguchi}, {Saito}, {Mikami}, {Hill}, \&
  {Owens}}]{Hinode5}
{Tsuneta}, S., {Ichimoto}, K., {Katsukawa}, Y., {et~al.} 2008, \solphys, 249,
  167

\bibitem[{{van Noort}(2012)}]{2012vannoort}
{van Noort}, M. 2012, \aap, 548, A5

\bibitem[{{van Noort} {et~al.}(2005){van Noort}, {Rouppe van der Voort}, \&
  {L{\"o}fdahl}}]{2005vannoort}
{van Noort}, M., {Rouppe van der Voort}, L., \& {L{\"o}fdahl}, M.~G. 2005,
  \solphys, 228, 191

\bibitem[{{van Noort} \& {Rouppe van der Voort}(2008)}]{2008vannoort}
{van Noort}, M.~J. \& {Rouppe van der Voort}, L.~H.~M. 2008, \aap, 489, 429

\bibitem[{{Wittmann}(1969)}]{W69}
{Wittmann}, A. 1969, \solphys, 7, 366

\end{thebibliography}

\end{document}